\documentclass[aip,reprint]{revtex4-2}
\usepackage[utf8]{inputenc}
\usepackage{graphicx}
\usepackage{dcolumn}
\usepackage{bm}
\usepackage{xcolor}
\usepackage{mathtools}
\newcommand{\p}{$\%$}

\newcommand{\pn}{$\mathrm{R{_{N_2}}}$}

\newcommand{\tcn}{$\mathrm{Co_{4}N}$}

\newcommand{\tfn}{$\mathrm{Fe_{4}N}$}
\newcommand{\Ts}{$\mathrm{T_{s}}$}
\newcommand{\Ta}{$\mathrm{T_{a}}$}
\newcommand{\Tc}{$\mathrm{T_{C}}$}

\begin{document}
\title{Synthesis and study of fcc-Co derived from isostructural \tcn}
\author {Seema, Dileep Kumar, U. P. Deshpande and Mukul Gupta}\email{mgupta@csr.res.in (M. Gupta)}
\address{UGC-DAE Consortium for Scientific Research, University Campus, Khandwa Road, Indore 452 001, India}

\date{\today}
\begin{abstract}
This work demonstrates synthesis and study of fcc-Co derived from an isostructural~\tcn. Diffusion measurements carried out in this work, reveal that N self-diffusion is the swiftest in \tcn~compared to other transition metal nitrides or even the mononitride CoN. By the application of a high substrate temperature (\Ts) growth or thermal annealing temperature (\Ta); N diffuses out from the fcc-\tcn~above 573\,K leaving behind a high purity fcc-Co phase. Generally, Co grows in a hcp structure and a (partial) hcp$\rightarrow$fcc-Co transformation takes place around 700\,K or above 70\,GPa. The proposed route through nitridation and diffusion of N not only bring down the phase transition temperature, an impurity present in the form of hcp-Co can be avoided altogether. Oriented \tcn(111) thin films were grown using a CrN(111) template on a quartz substrate using a dc magnetron sputtering. Samples were grown at different \Ts~or room temperature grown \tcn~samples were annealed at different \Ta. Analysis using x-ray diffraction, N K-edge x-ray absorption, x-ray photoelectron and secondary ion mass spectroscopy confirmed the formation of fcc-\tcn~or fcc-Co phases. It was found that Co-N bonding and N concentration get significantly reduced at a high~\Ts~or~\Ta. Magnetization measurements combining ex-situ and in-situ magneto-optical Kerr effect showed differences in anisotropy and coercivity of \tcn~and fcc-Co samples. Combining structural, electronic and magnetization measurements, it has been observed that a high purity fcc-Co can be conveniently derived from the isostructural~\tcn~aided by an exceptionally high N self-diffusion in~\tcn.

\end{abstract}

\maketitle

\section{Introduction}
Cobalt (Co) is well-known to be formed in three allotropes namely (i) anisotropic high-coercivity (H$_c$) hcp-Co (ii) pseudo-cubic bcc-Co and (iii) symmetric low H$_c$ fcc-Co. Generally, under ambient conditions, Co crystallizes in a hcp structure and transforms into fcc structure above 700\,K or 70\,GPa~\cite{2017_ScRep_Co,PRB:CoN:07}. Among the three allotropes, fcc-Co phase has been proposed to be used in soft magnetic applications involving power electronics and biomedicine. It has been reported that soft fcc-Co phase with high saturation magnetization and reasonably low H$_c$ is a good candidate for medical diagnostic and therapy applications~\cite{2007_JMMM_Co_application}. Recently, some theoretical as well as experimental study on the structural and magnetic transformations in bulk Co had been performed suggesting that as Co transforms from hcp to fcc phase at high pressure and the magnetization vanishes at 150 GPa~\cite{2019_PhysRevB_liq_Ni_Co_XAS,2016_PRB_Co_HPLT}. Various synthesis methods have been utilized to produce nanocrystalline fcc-Co with a suitable control of the particle size, stability, and magnetic properties. These methods involve chemical routes, solvothermal process, air synthesis as well as thin film deposition techniques~\cite{1998_hcpfccphasediagram,2006_GPSynthesis_Co,2009_MCP_Co_synthesis,GAJBHIYE2008Co,2007_JMMM_DK,2019_TSF_Banu_Co}. In a study reporting Fischer-Tropsch synthesis of Co particles, it was observed that hcp-Co phase partially exist even at 850\,K~\cite{2015_hcp_fcc_Co_NMR}. Recently, Kumar et al.~\cite{2019_kumar_Co_in_C} studied biphasic Co in C matrix using pyrolysis synthesis and found fcc-Co formation dominating at high heating rates. 

In thin film form, usually high temperature film deposition or annealing has been carried out for the hcp to fcc transformation~\cite{2007_JMMM_DK}. Even by high temperature annealing at $\approx$ 773\,K, partial transformation to fcc Co has been observed and hcp+fcc phase were found to coexist. Surprisingly, Banu et al.~\cite{2017_SR_Banu_HDNM_fccCo,2018_IOP_Banu_fccCo} reported the formation of high-density non-magnetic fcc-Co layer at the top in polycrystalline Co thin films under ambient conditions. Among thin films, mostly hcp(0001) or fcc(111) oriented films have been studied along with very few reports of meta stable bcc Co~\cite{2011_ohtake_Co}. In polycrystalline films,
generally stacking faults exist parallel to the closed packed plane of fcc and hcp structure and even small volume of crystals with mixed orientation exist~\cite{longo2014_Co,2021_JAC_Seema_Co4N}. However, once fcc-Co formed, reverse transformation from fcc to hcp phase is not favorable because of energetics involved in the process~\cite{fcc_hcp_theory}. 

Nitrides of Co have been recently explored extensively in search of materials for metal-air batteries in energy storage. Tetra cobalt nitride (\tcn) possess multiple magnetic functionalities and has been proposed as an interesting material for spintronic devices. They are also an active catalytic agent in oxygen evolution reaction and reduction reactions~\cite{kang2018}. The formation of ~\tcn~phase takes place when all the bcc sites in fcc-Co are occupied with N. The theoretical value of lattice parameter of \tcn~is about 3.7\,\AA~but the highest observed experimental value was 3.65\,\AA~\cite{JAC16_NPandey}, and it tends to decrease when \tcn~thin film samples were deposited at higher~\Ts or annealed at high~\Ta~\cite{JVSTA:Fang:CoN,CoN_AIP_Adv2015,JAC16_NPandey}. Literature reports on N self-diffusion measurements in mononitride phases of transition metals suggests that the N diffuses faster in CoN as compared to FeN and other early transition metal nitrides~\cite{PRB:AT:2014,PhysRevB.NP}. This trend has been observed to follow the order of enthalpy of formation of these compounds. Since~\tcn~has been predicted to have a similar enthalpy of formation ($\approx$~0\,kJ/mole~\cite{imai2014}) as that of CoN, it is anticipated that N-diffusion might inhibit the formation~\tcn~phase at higher temperatures. Also, a recent study focused on the process of N incorporation in Co lattice and found it minimal at temperature more than 623\,K using NH$_3$~plasma~\cite{2020_acs_jpcc_CoNx}. Therefore, this phenomenon of N self-diffusion at higher temperatures can be utilized for obtaining fcc-Co form~\tcn~anticipating a complete out diffusion of N. 

Therefore, in this work, we studied the phase formation of ~\tcn~at different~\Ts~and characterized resulting samples. For comparison purpose, we also annealed~\Ts~=300\,K~deposited~\tcn~samples at different temperatures and compared them with those deposited at different \Ts. We characterized crystal and electronic structure as well as the magnetic properties of deposited and annealed samples. We also quantified N self-diffusion in~\tcn~phase for the first time. This work provides an alternate pathway for the formation fcc-Co phase derived from the isostructural~\tcn. 

\section{Experimental}
Polycrystalline Co-N thin films were deposited on an amorphous
quartz (SiO$_2$) substrate using a direct current magnetron sputtering (dcMS) system
(Orion-8, AJA Int. Inc.). A high purity $\phi$~3\,inch Co target (99.995\p)
was sputtered with a dc power of 180\,W.
During deposition an Ar+N$_2$ gas mixture was used where
total flow was 50\,sccm and N$_2$ gas flow (\pn) was 10\,sccm. This \pn~value is an optimized one for the phase formation of~\tcn~\cite{2021_JAC_Seema_Co4N}. The base pressure of the chamber was
better than 5$\times$10$^{-8}$\,hPa, while during deposition
the working pressure was about 5$\times$10$^{-3}$\,hPa. Substrate temperature (\Ts) was varied from 300 to 673\,K in steps of 100\,K. All samples were deposited on a 50\,nm CrN buffer layer to prevent film-substrate intermixing that might take place at high~\Ts.
Typical thickness was about 200\,nm for all samples. For N-diffusion measurements, special trilayer structured film was deposited on Si (100) substrates using dcMS. Details regarding this sample are explained in section~\ref{section:diffusion}.

After deposition, small pieces were cut from 300\,K deposited film and those were vacuum annealed separately at the different~\Ta~for 1\,hour. Annealing was carried out in a homemade set up in which a base the pressure of 1$\times$10$^{-6}$\,hPa was achieved before annealing. During annealing, the pressure was 5$\times$10$^{-6}$\,hPa. Crystalline structure were characterized by x-ray diffraction (XRD) using a standard x-ray diffractometer
(Bruker-D8 Advance) using Cu K$_{\alpha}$ x-rays in $\theta$-2$\theta$ geometry. Chemical depth profile of the as-deposited and annealed samples were measured using secondary ion mass spectroscopy (SIMS) using O$_2^+$ as source to sputter out the film with 5\,keV energy and 400\,nA beam current. Local structure of the samples were analyzed by soft x-ray absorption spectroscopy (XAS) at BL-01 beamline (Indus-2) in total electron yield mode~\cite{XAS_beamline}. X-ray photoelectron spectroscopy (XPS) measurements were carried out using Mg K$_{\alpha}$ x-ray source after Ar ion sputtering of the top few nm of the sample. Effect of \Ts~and \Ta~on the magnetic properties were studied using magneto-optic Kerr effect (MOKE) in longitudinal geometry with Evico magnetics setup. Further, temperature dependent MOKE measurements were also carried out with simultaneous heating and in-situ hysteresis measurement in a UHV chamber.

\section{Results}

\subsection{Effect of \Ts~or~\Ta~on structure}

\begin{figure*} [ht]
	\centering
	\includegraphics [width=140mm] {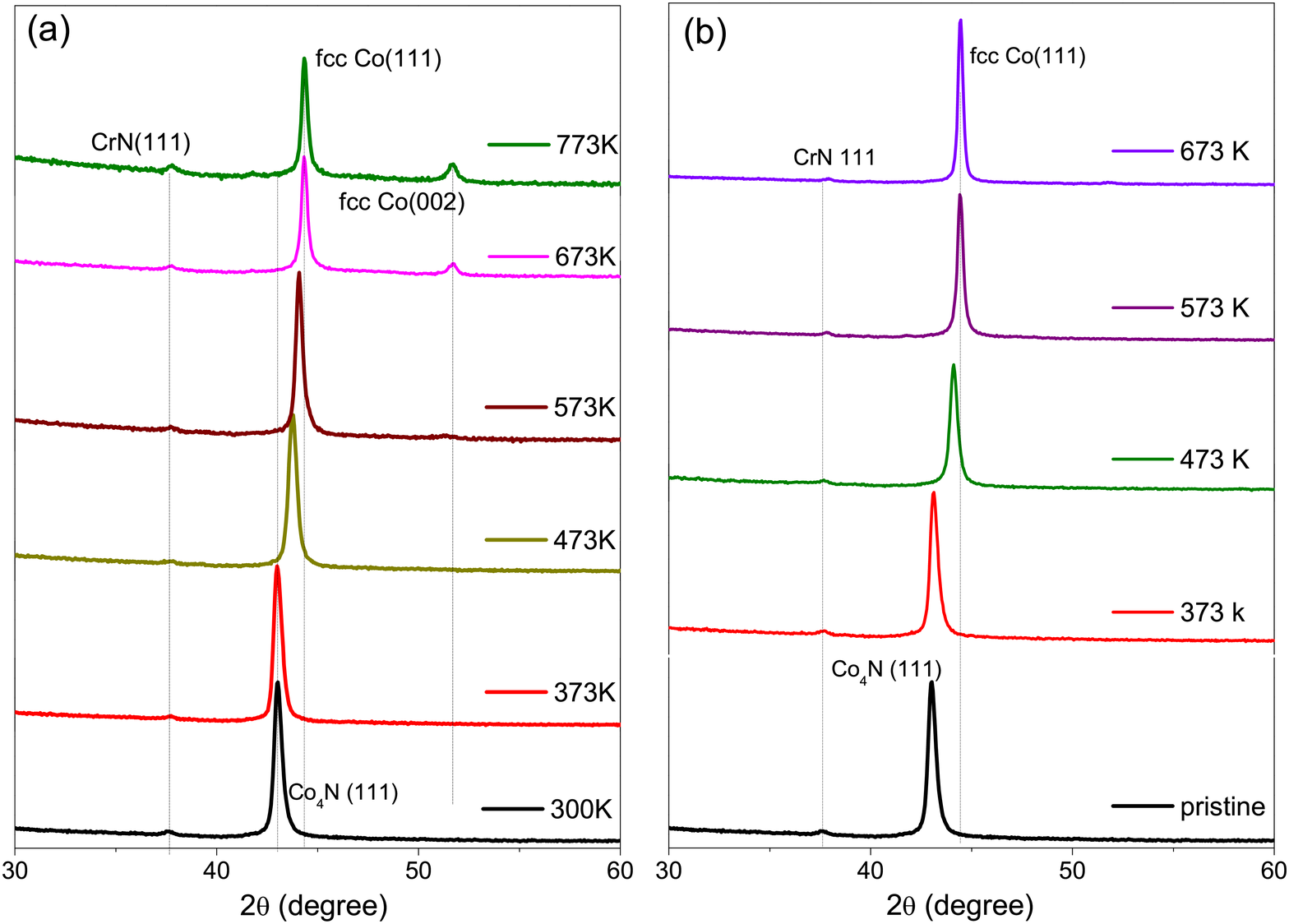}
	\caption{\label{fig:xrd_all} (a) XRD data of \tcn~thin films
		deposited at different substrate temperature (\Ts). (b) XRD patterns of \tcn~thin film deposited at 300\,K	annealed at different temperatures \Ta.}
	\vspace{-1mm}
\end{figure*}

\begin{figure} [ht]
	\centering
	 \vspace{-5mm}
	\includegraphics [width=85mm] {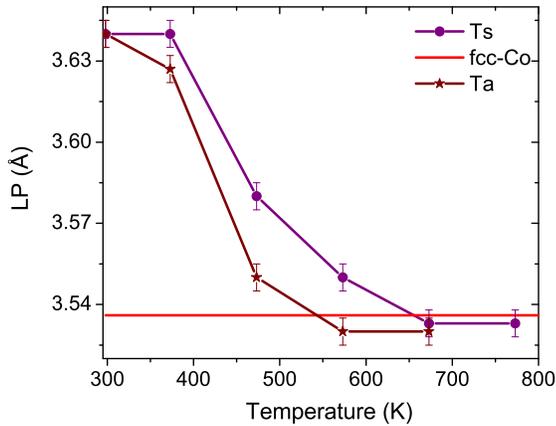}
	\caption{\label{fig:lp} Comparison of lattice parameter (LP) of \tcn~films with different \Ts~and \Ta.}
\end{figure}

XRD data of samples deposited at different~\Ts~are shown in fig.\ref{fig:xrd_all} (a). In the \Ts~=~300\,K sample peaks corresponding to CrN (111) and \tcn~(111) can be seen. The lattice parameter (LP) of~\tcn~phase comes out to be 3.645~($\pm$0.005)\,\AA, in agreement with previous works~\cite{CoN_AIP_Adv2015,2001_Vac_Asahara}. The notable difference here is the absence of impurity peaks that are generally seen in Co-N compounds [REF].  It can be anticipated that the uasge of CrN (111) templet favors the growth of~\tcn~(111). When the \Ts~is raised to 373\,K, XRD pattern remains similar to the one observed for Ts~=~300\,K, but when the \Ts~was raised to 473\,K and beyond, the peak keeps on shifting towards higher 2$\theta$. However, beyond 673\,K it becomes stagnant. A presence of faint fcc Co (200) reflection can also be seen in Ts~$\geq$~573\,K samples. From the observed behavior, it can be inferred that samples grown at Ts~=~300 and 373\,K have LP~=~3.645~($\pm$0.005)\,\AA but it reduces to 3.532~($\pm$0.005)\,\AA (a value expected for fcc Co) when the Ts~exceeds 673\,K. The variation of LP with the \Ts~is shown in fig.\ref{fig:lp}. These LP values were calculated after fitting the XRD peak profile to a single Lorentzian function~\cite{Scherrer_Eq:NatureNano:2011,JMMM_MG}.

The \Ts~=~300\,K sample was annealed at similar temperature (\Ta~=~\Ts) and observed XRD patterns are plotted in fig.\ref{fig:xrd_all} (b). Here also a similar behavior can be seen albeit a much steeper fall in the LP can be observed and the absence of Co(002) reflection can be notices as shown in fig.\ref{fig:lp}. Contrary to the decomposition of~\tcn~into Co$_3$N and then to fcc-Co~\cite{CoN_AIP_Adv2015,RTA_CoxN}, we observed a straight away decomposition into fcc Co probably due to much larger steps taken in the~\Ta. Similar observation of phase decomposition had also been observed for \tfn~phase~\cite{2019_physB_Fe4N_NP}. Our XRD results clearly indicate that N seems to completely diffuse out from~\tcn~when the~\Ts~$\geq$~673\,K or the \Ta~$\geq$~573\,K. To further confirm such claims, elemental depth profiles were measured and presented in the next section.

\begin{figure} [ht]
	\centering
	\includegraphics [width=85mm] {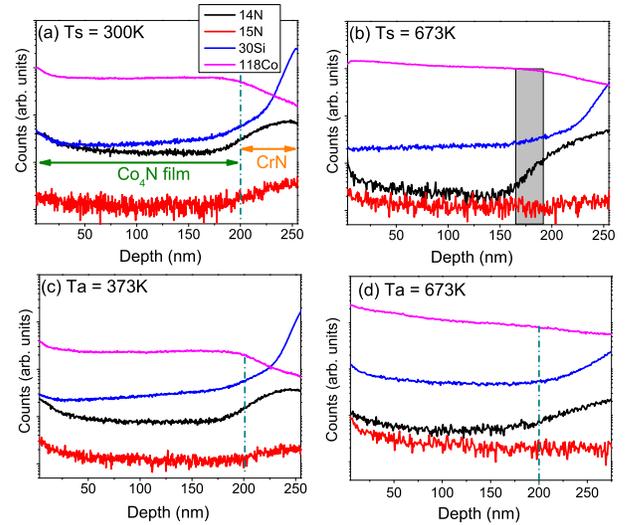}
	\caption{\label{fig:sims} Representative SIMS depth profile of 14N (black), 15N (red), 30Si (blue) and 118Co (magneta) in \tcn~films deposited at different \Ts~=~300\,K (a) and 673\,K (b), and annealed at different \Ta~=~373\,K (c) and 673\,K (d). The vertical line is guide to the eye indicating \tcn~film-CrN buffer layer interface.}
	\vspace{-1mm}
\end{figure}

\subsection{Effect of \Ts~or~\Ta~on elemental depth profiles}

Fig.~\ref{fig:sims} shows the SIMS depth profiles of $^{14}$N, $^{15}$N, $^{30}$Si and dimer $^{118}$Co in samples deposited or annealed at different~\Ts~or~\Ta. Since the abundance of $^{15}$N is only about 0.4\p~in N$_2$ therefore its counts are expected to be much lower as compared to those of $^{14}$N. This is indeed the case for N profiles in CrN buffer, but in case of \tcn~samples such difference can only be seen in case of Ts~=300\,K (fig.~\ref{fig:sims}a) or \Ta~373\,K (fig.~\ref{fig:sims}c) samples. As N diffuses out from \tcn~we can see that count of $^{14}$N become considerably low and in fact they start to merge together with $^{15}$N both in case of \Ts~=~673\,K samples (fig.~\ref{fig:sims}b) or \Ta~=~673\,K (fig.~\ref{fig:sims}d). This is a clear indication that indeed N diffuses out from~\tcn~at high~\Ts~or~\Ta. In addition, it can be seen that Co profiles are uniform in low~\Ts~or\Ta~samples but a clear gradient can be seen in~\Ts~or\Ta~=~673\,K samples indicating Co-richer regions created by N migration from the surface. The $^{30}$Si profiles are indicative of reaching the substrate due to a substantial rise in its count. In general our SIMS depth profile results correlate well with the XRD analysis. 

\subsection{Effect of \Ts~or~\Ta~on electronic structure}

\begin{figure*} \centering
	\includegraphics [width=120mm] {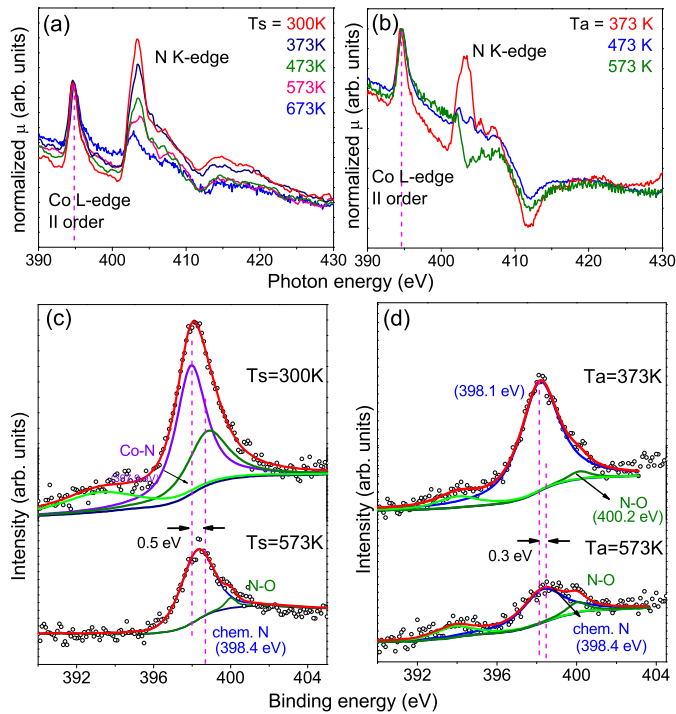}
		\caption{\label{fig:xas_xps} N K-edge spectra of \tcn~films with different \Ts~(a) and \Ta~(b). N 1s spectra pristine \tcn~sample deposited at \Ts~=~300\,K and comparison with sample deposited at \Ts~=~573\,K and (d) N 1s spectra when pristine \tcn~annealed at \Ta~=~373 and 573\,K.}
\end{figure*}

The electronic structure measurement of samples was carried out combining N K-edge XAS and N 1s and Co 2p XPS measurements. The N K-edge XAS shown in fig.~\ref{fig:xas_xps} (a),(b) reveals prominent features around 394 and 403\,eV. In which the feature around 394\,eV arises due an experimental artifact caused by second order diffraction of Co L-edge from the grating monochromator. The sharp feature at 403\,eV corresponded to N K-edge and arises due to dipole transition coming from N 1s to hybridized states to fcc Co 3d and N 2p orbitals through $\pi$$^{\star}$ anti-bonding~\cite{2019NP_Co_1_3inch}. As the ~\Ts~or~\Ta~increases, though the intensity of Co L-edge feature remains intact, that of N K-edge gradually falls and reaches to the background at the highest \Ts~or~\Ta. This again indicates that N is diffusing out from~\tcn~at high~\Ts~or~\Ta~in agreement with the observations made from XRD and SIMS results. 

Representative N 1s spectra can be seen in fig.~\ref{fig:xas_xps} (c),(d). They were fitted using XPSPEAK41 software. The most intense feature around 397.9\,eV in fig.~\ref{fig:xas_xps} (c) can be attributed to the Co face centered atoms bonded with N~\cite{yao2019CrCo4N,kim2006CoSi2}. This feature is quite distinct in~\Ts~=~300\,K sample due to the formation of~\tcn~phase. In the~\Ts~=~573\,K sample, this feature not only reduces in intensity it as well shifts to higher binding energy by 0.5\,eV. The loss of intensity is an indication of lesser N atoms and a change in the bonding state of Co atoms~\cite{2006_Cruz_CoN} can also led to similar effects. Along with this, an additional unidentified feature is present in both these samples around B. E. =  393\,eV. In 300\,K sample, the feature around 399\,eV corresponds to interstitially trapped or chemisorbed species on N, while in~\Ts~=~573\,K sample, the feature at 400.1\,eV depicts some N-O bonds present, however the relative area of these features is small. In fig.~\ref{fig:xas_xps} (d) the N 1s spectra of the annealed samples is shown. From a direct observation, it can be inferred that the N 1s spectra of annealed sample is different from~\tcn~sample prepared at different~\Ts~(as in fig.\ref{fig:xas_xps} (d)). The main feature for 373\,K annealed sample is Co-N bonding (398\,eV)~\cite{2018CoCo4NMAB,fan2019ALD,2020_AFM_TiNCo4N} and some N-O bonding (400\,eV). For sample annealed at 573\,K, the intensity of N 1s spectra is quite low as compared to others. To further quantify this, we compared the background subtracted area of the N 1s peak of samples. As the~\Ts~is raised to 573\,K, the area of N 1s peak gets reduced by 75\p. For~\Ta~=~373 and 573\,K samples, the N 1s peak area reduced by 50 and 80\p, respectively as compared to~\Ts~=~300\,K depsoited~\tcn~sample. This small concentration of N may be attributed to some N trapped at interstitial sites due to N self-diffusion (discussed in section~\ref{section:diffusion}). Therefore, it can be concluded that temperature (\Ts~as well as~\Ta) affects the Co-N bonding up to large extent as N 1s peak area gets significantly reduced at higher~\Ts~or~\Ta.

\begin{figure} [ht]
	\centering
	\vspace{5mm}
	\includegraphics [width=105mm] {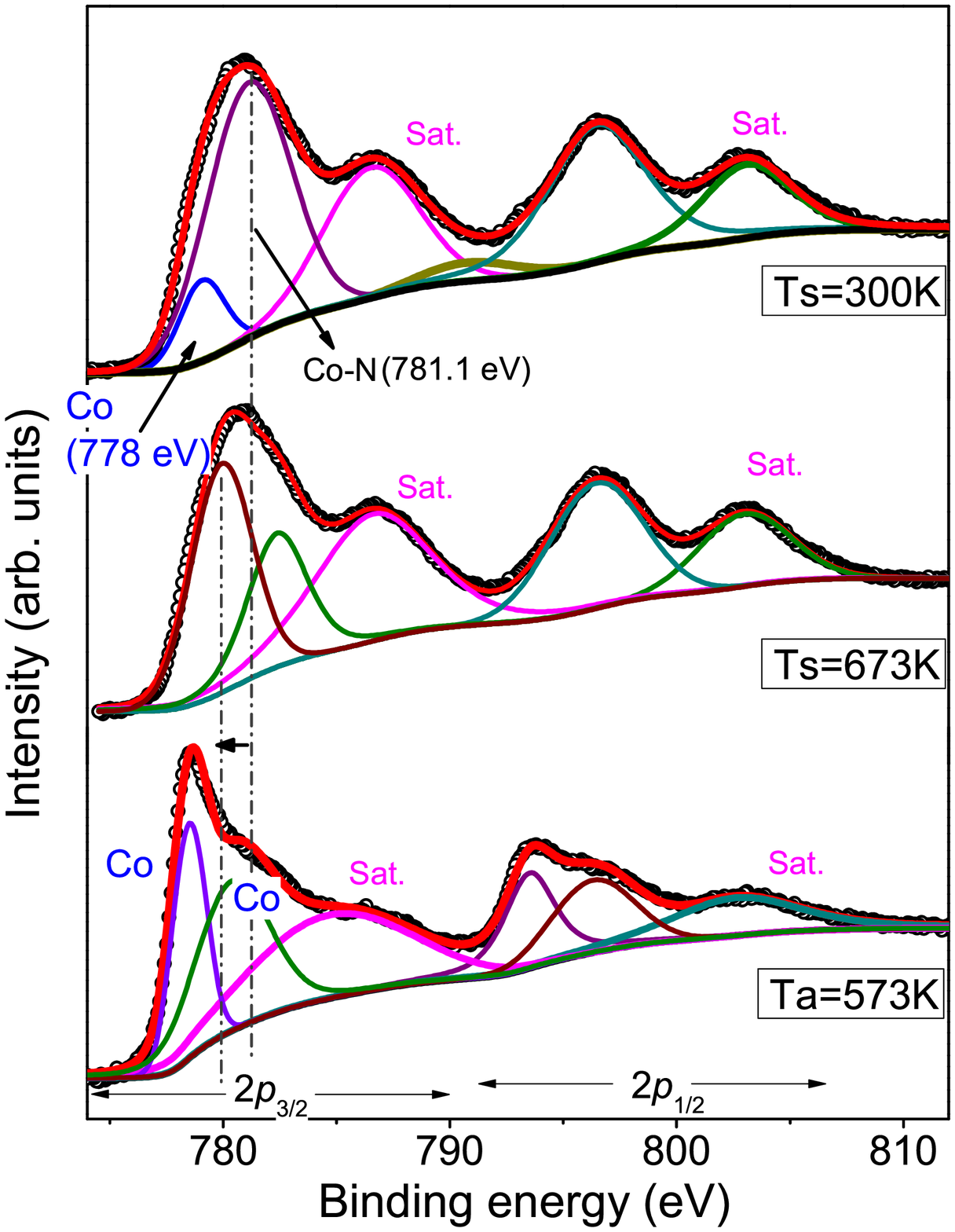}
	\caption{\label{fig:xps_Co} Co 2p XPS spectra with temperature variation.}
\end{figure}

Similar results were reflected in XPS measurements on Co 2p spectra of these samples and the representative spectra are shown in fig.\ref{fig:xps_Co}. The distinct features of these spectra are Co-Co (778.2\,eV), Co-N (780.6\,eV) in both Co 2p$_{3/2}$ and 2p$_{1/2}$ regions due to spin-orbit splitting~\cite{kang2018,fan2019ALD,Co4N_CNT,2007_Yao_Co4N_Al2O3}. The presence of dominant Co-N binding feature at 300\,K again confirms \tcn~phase formation and the oxidation state of Co for \tcn~between +2 and +3~\cite{2007_Yao_Co4N_Al2O3}. The Co-N feature shifts and diminishes as \Ts~is increased to 573\,K, which again corroborates with our previous results of \tcn~phase decomposition. However, the presence of additional shifted feature in \Ts~=~673\,K sample might arise due to N atmosphere during film deposition. In 573\,K annealed sample, the spectra is quite different as it is shifted to lower B. E., which shows change in chemical state of Co. This spectra was fitted with different components which could be due to Co-Co bond and satellite peaks which generally lie 3-5\,eV above main peak. These observations are in accordance with the N 1s spectra of these samples and again proves the fact that N incorporation is less at high \Ts~and \Ta.

\subsection{Effect of \Ts~or~\Ta~on magnetic properties}

\begin{figure*} \centering
	\includegraphics [width=170mm] {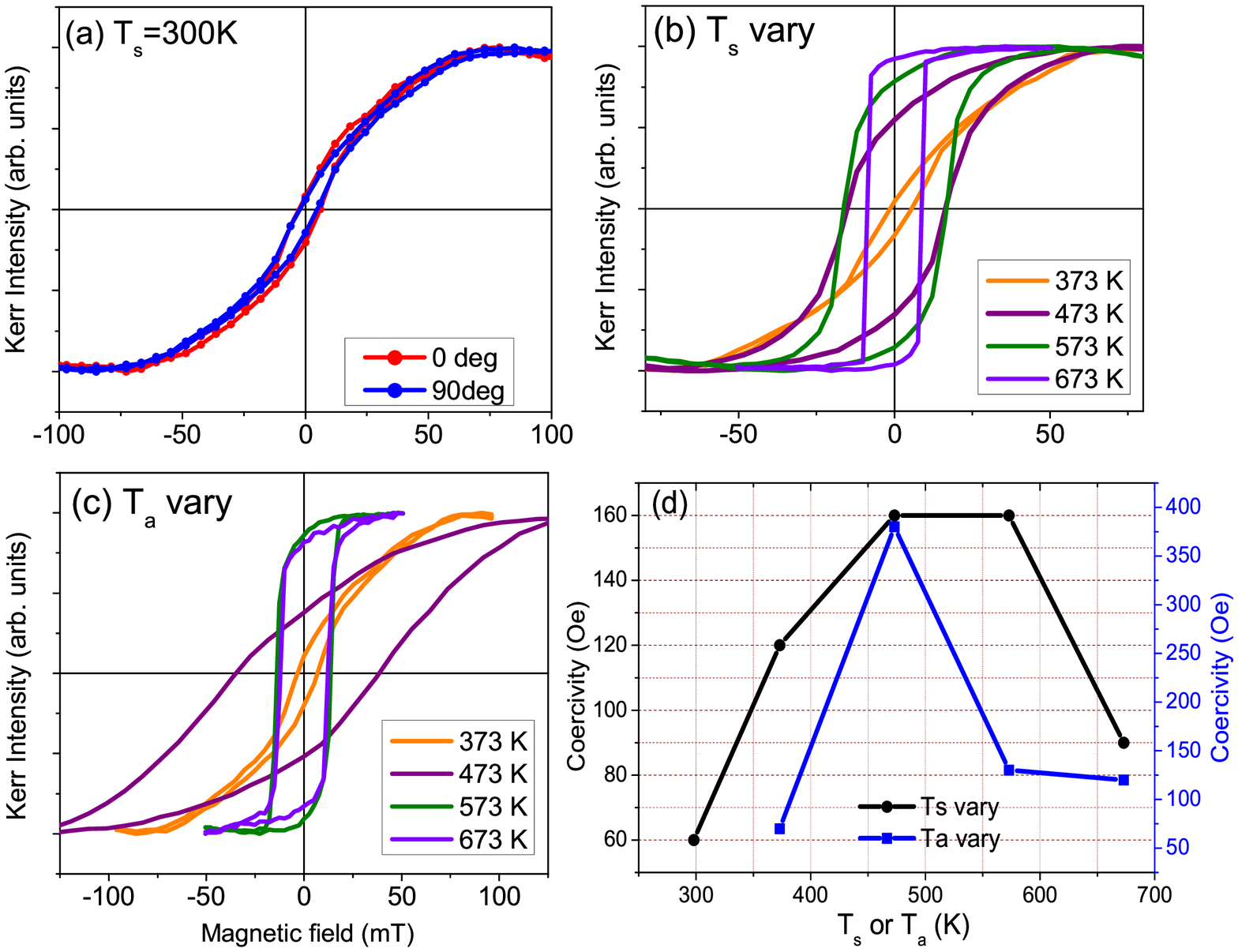}
	\vspace{-10mm}
	\caption{\label{fig:moke_all} L-MOKE hysteresis of \tcn~film deposited at 300\,K, with in-plane rotation angles (a). Hysteresis loop for different \Ts~(b) and \Ta~(c). The variations in Coercivity with \Ts~or\Ta~are shown in (d) and the errors in the value of Coercivity is about the same as the size of symbols.}
	\vspace{-2mm}
\end{figure*}

The saturation magnetization behavior of Co-N films deposited at different~\Ts~was studied by Li et al.~\cite{li2017CoN}. In the present work, we studied the magnetic anisotropy (MA) using L-MOKE as shown in fig.\ref{fig:moke_all}. It is known that for a given applied field along the easy axis of magnetization, the loop show a perfect square shape. However, for~\tcn~film deposited at~\Ts~=~300\,K, the hysteresis loop is round shaped (fig.\ref{fig:moke_all} (a)). Such rounding-off of a loop indicate that the saturation is taking place by domain rotation with the applied field. This could happen due to the internal stresses present in the film at the time of deposition itself~\cite{2007_JMMM_DK}, as observed previously in case of Co thin films. It had been reported that, for the case of~\tfn~films that saturation in [111] direction is difficult, which could also lead to such shape of the loop. Also, this sample is magnetically isotropic as no changes were observed when the sample is rotated in the plane of applied magnetic field~\cite{2000annealedCo_MA}. 

For~\Ts~=~373\,K and~\Ta~=~373\,K samples, the obtained hysteresis loops are identical. But the saturation behavior as well as the coercivity (Hc) of the sample changed above 473\,K. In~\Ts~=~473\,K sample, the MA behavior is quite different, and an enhancement in the Hc is observed as shown in fig.\ref{fig:moke_all} (b) and fig.\ref{fig:moke_all} (d). Also, the remnant magnetization and squareness increases. This can be correlated with the local structure changes taking place in the sample which have been observed in from various results presented in this work. On further higher~\Ts, the squareness of the loop increases, while the Hc remains the same. For~\Ts~=~673\,K, the Hc get reduced and the squareness approaches to 1. Such a variation can be correlated with the phase change from~\tcn~to fcc-Co. For annealed sample, the squareness and the Hc variation is somewhat different. Here, for sample annealed at 473\,K, the saturation field is high. Although, the remnant magnetization is small as compared to~\Ts~=~473\,K sample, the Hc for both are observed to be almost similar. This variation of Hc with~\Ts~and~\Ta~is shown in fig.\ref{fig:moke_all} (d). The sudden increase in the Hc is also a signature of fcc-Co formation as reported by Kumar et al.~\cite{1996_APL_Co-N_matsouka,2007_JMMM_DK}. However, our fcc-Co samples show overall ferromagnetic behavior contrary to the non-magnetic fcc-Co dense layer observations by Banu et al.~\cite{2017_SR_Banu_HDNM_fccCo}. On further annealing of \tcn~sample at 573 and 673\,K, the loops are almost square with high remnant magnetization and low Hc. The changes observed in the Hc and hysteresis behavior can be correlated with the fact that during both \Ts~and \Ta~variation, phase changes are taking place. On the one hand, the  chemical change due to \Ts~occur at the time of film deposition itself, the homogeneity of Co and N along the depth of film remains intact till 573\,K. On the other hand, annealing process disturbs the chemical homogeneity of the film as observed in our SIMS depth profiles. Due to difference in nature of these two processes, the MA and Hc variations are different individually, however the overall behavior is almost similar.

\begin{figure} [ht] \centering
	\includegraphics [width=95mm] {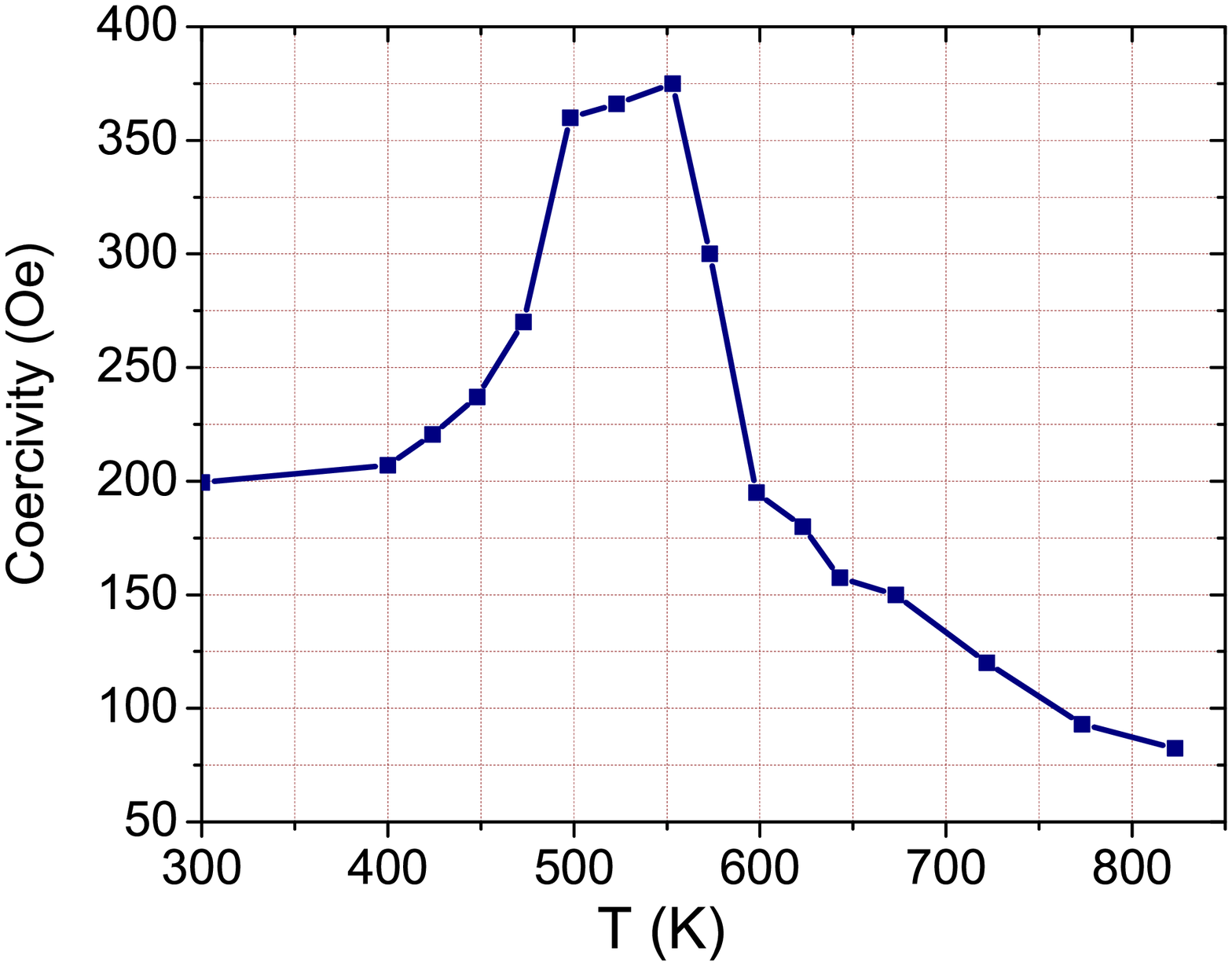}
	\vspace{-21mm}
	\caption{\label{fig:insitu_moke} (a) L-MOKE hysteresis of \tcn~film deposited at 300\,K and measured in-situ in a UHV system at different annealing temperatures. Errors in estimation of Coercivity is typically the size of symbols.}
	\vspace{-7mm}
\end{figure}

To further clarify the dependent of variation in Hc with \Ta, we performed in-situ temperature dependent MOKE measurements under UHV conditions. For these, sample is mounted on a holder which is transported via load lock on to a heating stage situated inside UHV chamber with L-MOKE arrangement~\cite{JAP_AT_FeN_Al_Zr}. The temperature was increased at a steady rate of 1\,K/min and hysteresis loops were measured continuously. The variation of Hc of \tcn~sample with temperature is shown in fig.~\ref{fig:insitu_moke}. Here, the observed behavior is similar to the \Ts~or \Ta~variation and the region in which the variation in Hc is highest between 475 to 575\,K. As the temperature is raised at a steady rate, the Hc of sample first increases, attains a plateau and then decreases. The variation of Hc can be correlated with phase transormation from \tcn~to fcc-Co+\tcn~to fcc-Co. This observed behavior matches well with the report by Matsouka et al.~\cite{1996_APL_Co-N_matsouka}. These observations are in agreement with the ex-situ MOKE results of \Ts~and \Ta~variation.

\subsection{N self-diffusion measurements}
\label{section:diffusion}

\begin{figure*} \centering
	\includegraphics [width=130mm] {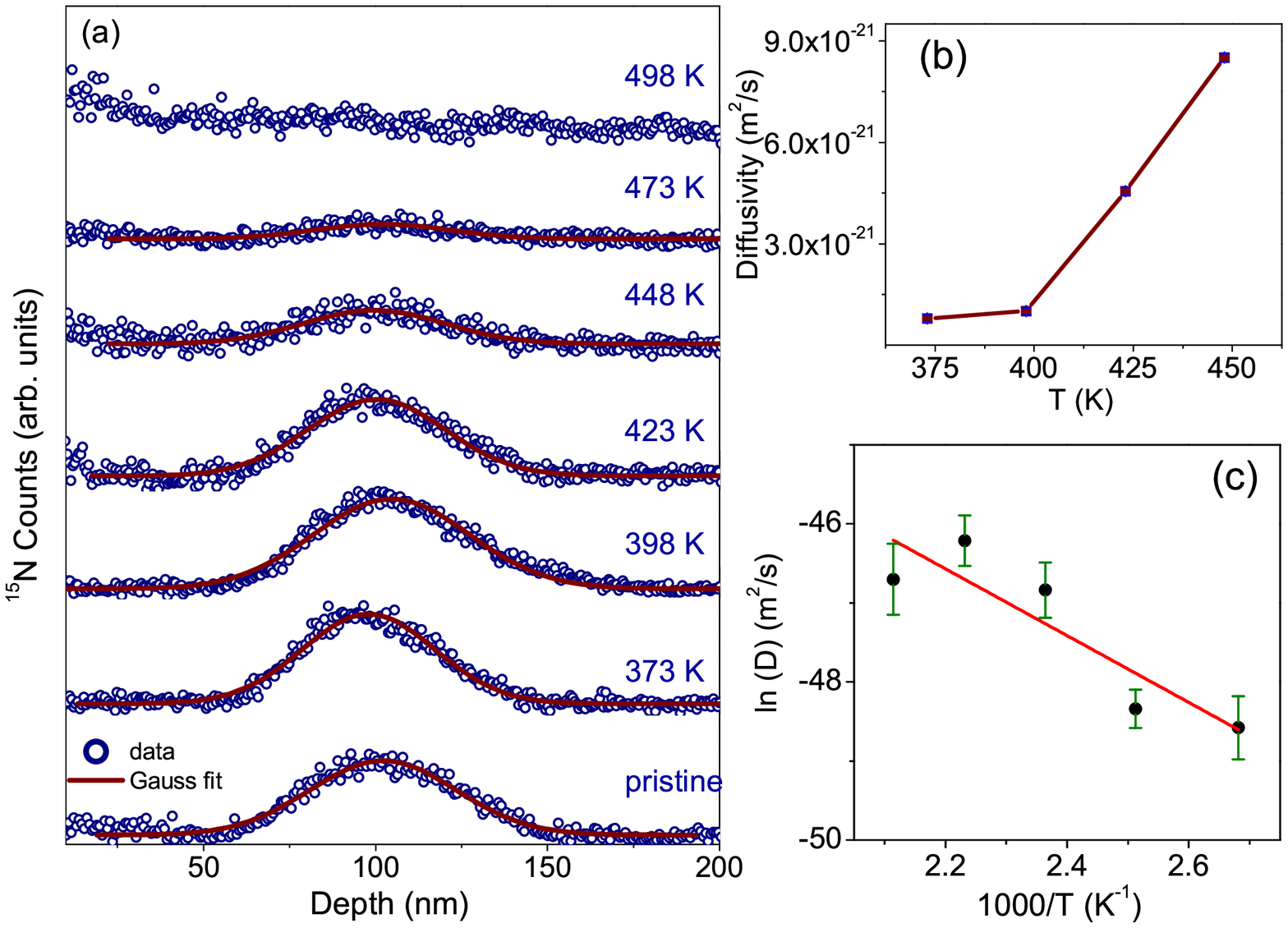}
	\caption{\label{fig:diffusion} (a) SIMS depth profile of $^{15}$N in Co-N samples on Si (100) substrate vacuum annealed at different temperatures separately, (b) Diffusivity (D) as a function of annealing temperature. The error bars are size of the symbol. (c) Arrhenius plot of diffusivity (D) of N.}
\end{figure*}

Since the arguments related to the effect of the~\Ts~or~\Ta~on the structural, electronic and magnetic properties of samples have been explained in terms of out diffusion of N from~\tcn, we measured N self-diffusion in~\tcn~adopting a similar approach that was made in case of mononitride  CoN~\cite{PhysRevB.NP}. To do so, a $^{15}$N enriched layer was sandwiched between their natural counterparts on a Si substrate as: [\tcn(100\,nm)$\mid$Co$_4^{15}$N(5\,nm)$\mid$\tcn(100\,nm)]. This sample was deposited at~\Ts~= 300\,K. The presence of$^{15}$N was ensured using a residual gas analyzer during the deposition process. It is expected that the SIMS depth profile of $^{15}$N in this sample will show a peak due to its higher concentration. Also, when this sample is annealed at different temperatures, the broadening of the peak will provide information about self-diffusion of N in \tcn~phase. Such measurements have been performed in mononitride FeN and CoN~\cite{PRB:AT:2014,PhysRevB.NP}, but not in~\tcn. The diffusion analysis is based on Fick's second law, according to which the diffusing particles (here $^{15}$N) get dispersed into two material bodies with a constant and equal diffusivity~\cite{Shewmon}.

$^{15}$N SIMS depth profiles of
the as-deposited film and vacuum annealed samples (isochronally for 1\,h) at different
temperatures are shown in fig.~\ref{fig:diffusion} (a). We can clearly see from
this figure that by increasing the annealing temperature, the $^{15}$N peak profile broadens. The width of this $^{15}$N peak at different temperatures can be used to calculate the time average diffusivity
($D$) as~\cite{MG_chap}:

\begin{equation} \centering
	\label{eq:av_D}
	D (t) = \frac{\sigma^2_t-\sigma^2_0}{2t} .
\end{equation}

where $\sigma_0$, $\sigma_t$ are the standard deviation before annealing, $t$
= 0 and after an annealing time $t$~=~1 hour. Therefore, the obtained data is fitted with Gauss function to calculate $\sigma_t$, $\sigma_0$ and then $D$. $D$ as a function of temperature is plotted in fig.~\ref{fig:diffusion} (b), which shows a steep increment above 400\,K. Obtained $D$ of $^{15}$N
has been measured at different temperatures and plotted as ln($D$) vs. 1000/$T$. This variation is found to follow an Arrhenius type behavior, which can be used to estimate the activation energy $E$ using:

\begin{equation}\label{arh}
	D = D_0~\mathrm{exp}(-E/k_{B}T)
\end{equation}

where, $D_0$ denotes the pre-exponential factor, $k_B$ is Boltzmann constant and $T$ is annealing temperature, Using eq.~\ref{arh}, the values of $D$ obtained at different annealing
temperatures were fitted to a straight line as shown in fig.~\ref{fig:diffusion} (c). The slope of this line yields the activation energy ($E$) = 0.36$\pm$0.05\,eV
and $D_0$~=~2.5$\pm$1$\times$10$^{-17}$\,$\mathrm{m^2sec^{-1}}$. The value of $E$ 
obtained in \tcn~here is less than the value found in FeN, CoN, and Ni-N~\cite{PRB:AT:2014,PhysRevB.NP,MG_chap,2020_RRL_NiN_NP}. If we extrapolate the line in fig.~\ref{fig:diffusion} (c) to T~=~300\,K, we can estimate the diffusivity of N in \tcn~at room temperature. This value comes out to be 5$\times$10$^{-23}$\,$\mathrm{m^2sec^{-1}}$, even considering a typical error of 30\p, $D$ in~\tcn~is an order of magnitude higher as compared to mononitride CoN (4.9$\times$10$^{-24}$\,$\mathrm{m^2sec^{-1}}$)~\cite{PhysRevB.NP} and several orders of magnitude higher than other nitrides e.g. TiN (1.7$\times$10$^{-49}$\,$\mathrm{m^2sec^{-1}}$), CrN (1.0$\times$10$^{-55}$\,$\mathrm{m^2sec^{-1}}$) and FeN (1.3$\times$10$^{-33}$\,$\mathrm{m^2sec^{-1}}$)~\cite{PRB:AT:2014,PhysRevB.NP}. The high value of room-temperature $D$ in \tcn~suggest that N self-diffusion will be significant even for a sample which is kept at room temperature for a long period of time~\cite{PRB:AT:2014}. Estimating a time dependence on $D$ in one year (365 days), one would get a value which is 10 orders higher than $D$ of N in Fe-N phase as calculated by Tayal et al.~\cite{PRB:AT:2014}. This implies that N self-diffusion is significant even at ambient conditions in \tcn~phase. Therefore, it can be concluded that self-diffusion of N in \tcn~phase is higher  as compared to mono-nitride phases, which in turn depicts the poor thermal stability of \tcn. The higher value of $\Delta$$H^{\circ}_f$ makes N incorporation in Co lattice difficult and low value of $E$ for N make it diffuse out as soon as sample is heated. Overall, this clearly emphasize the poor thermal stability of the \tcn~system, and it can provide pathways for the formation of fcc-Co formation, which is otherwise very difficult to synthesize.

\section{Discussion} 
From the detailed study carried out in this work by measuring the structure, depth profiles, N K-edge XAS and N 1s XPS on samples deposited at different \Ts~or annealed at different \Ta~it becomes clear that N diffuses out from \tcn~$\approx$~600\,K and the resulting sample is fcc-Co. N self-diffusion measurements further confirm that it is exceptionally high in \tcn~as compared to other metal nitrides or even the mononitride CoN. From these measurements, a mechanism leading to the formation of fcc-Co from the isostructural \tcn~can be understood. Under ambient conditions, Co crystallizes in a hcp structure but at high pressure (70\,GPa) or high temperature (700\,K) it transform into fcc-Co albeit some amount of hcp-Co still remains. The route followed in this work is to first synthesize fcc-\tcn~via a reactive nitrogen sputtering of hcp-Co and thereafter N out-diffusion (by \Ts~or \Ta) from \tcn~leads to precipitation of fcc-Co above 573\,K. Once formed, the reverse transformation from fcc-Co to hcp-Co is not possible due to energetically unfavorable conditions~\cite{fcc_hcp_theory}. Though such transformation of fcc-Co from the isostructural \tcn~appears convincingly simplistic, it is to be noted that the Curie temperature (\Tc) of Co is quite high at 1400\,K. Since the transformation is taking place well below the \Tc, the local moments of Co are expected to play a deterministic role similar to those leading to hcp $\rightarrow$ fcc transformations in Co~\cite{2017_ScRep_Co}. Comparing the observed behavior with a similar compound \tfn, it appears that upon high temperature (above 800\,K) annealing of fcc-\tfn~converts it into bcc-Fe~\cite{chatbi1997nitrogen} instead of fcc-Fe. Here, it is to be noted that the \Tc~of Fe is much lower at around 1000\,K and the transformation from fcc-\tfn $\rightarrow$ bcc-Fe is also taking place near around the \Tc. Therefore, the role of local moments will be entirely different as compared to that in Co~\cite{fcc_hcp_theory}. Another effect that seems to be playing a deterministic role is the self-diffusion of N in these compounds. Among late transition metal nitrides (viz. FeN, CoN, Ni-N), N self-diffusion in CoN was found to be fastest~\cite{2020_RRL_NiN_NP} and in case of \tcn~N self-diffusion was even faster than the mononitride CoN. In this scenario, it becomes favorable for fcc-\tcn~to transform into fcc-Co at a moderate temperature of about 600\,K.             

\section*{Conclusion}
Through this work, we demonstrate synthesis of (111) oriented \tcn~films on a buffer layer of CrN (111) at a substrate temperature, \Ts~of 300\,K. Keeping all deposition conditions identical to the growth of \tcn~but at \Ts~$>$~573\,K resulting films resembles to fcc-Co (instead of \tcn). Also, thermal annealing of \tcn~film results into a similar transformation from \tcn~$\rightarrow$~Co when the annealing temperature \Ta, exceeds 573\,K. These observations reveal that at high enough \Ts~or \Ta, N diffuses out from fcc-\tcn, leaving behind the isostructural Co. Chemical depth profiling using SIMS and local structural and bonding analysis using XAS and XPS showed almost negligible N content in samples with \Ts~or \Ta~higher than 573\,K. Systematic changes in the magnetic properties of samples can be seen with the application of \Ts~or~Ta. Though the variation in magnetic anisotropy are quite similar in both cases, the magnitude of coercivity was found to be higher in annealed samples. N self-diffusion measurements in \tcn~phase were performed for the very first time and it was found that the diffusivity of N in \tcn~is exceptionally high as compared to neighboring transition metal or even with cobalt mononitride phase. Such a high diffusivity of N provides an alternative pathway for the formation of fcc-Co from the isostructural~\tcn~phase at a moderate temperature of about 600\,K.

\section*{Acknowledgments}

Cooperation received from R. Raj, S. Kaushik in MOKE measurements; P. Goyal in XPS; R. Sah and S. Kalal in XAS; and L. Behera in various experiments is gratefully acknowledged. 

\section*{References}

\begin{thebibliography}{49}%
	\makeatletter
	\providecommand \@ifxundefined [1]{%
		\@ifx{#1\undefined}
	}%
	\providecommand \@ifnum [1]{%
		\ifnum #1\expandafter \@firstoftwo
		\else \expandafter \@secondoftwo
		\fi
	}%
	\providecommand \@ifx [1]{%
		\ifx #1\expandafter \@firstoftwo
		\else \expandafter \@secondoftwo
		\fi
	}%
	\providecommand \natexlab [1]{#1}%
	\providecommand \enquote  [1]{``#1''}%
	\providecommand \bibnamefont  [1]{#1}%
	\providecommand \bibfnamefont [1]{#1}%
	\providecommand \citenamefont [1]{#1}%
	\providecommand \href@noop [0]{\@secondoftwo}%
	\providecommand \href [0]{\begingroup \@sanitize@url \@href}%
	\providecommand \@href[1]{\@@startlink{#1}\@@href}%
	\providecommand \@@href[1]{\endgroup#1\@@endlink}%
	\providecommand \@sanitize@url [0]{\catcode `\\12\catcode `\$12\catcode
		`\&12\catcode `\#12\catcode `\^12\catcode `\_12\catcode `\%12\relax}%
	\providecommand \@@startlink[1]{}%
	\providecommand \@@endlink[0]{}%
	\providecommand \url  [0]{\begingroup\@sanitize@url \@url }%
	\providecommand \@url [1]{\endgroup\@href {#1}{\urlprefix }}%
	\providecommand \urlprefix  [0]{URL }%
	\providecommand \Eprint [0]{\href }%
	\providecommand \doibase [0]{https://doi.org/}%
	\providecommand \selectlanguage [0]{\@gobble}%
	\providecommand \bibinfo  [0]{\@secondoftwo}%
	\providecommand \bibfield  [0]{\@secondoftwo}%
	\providecommand \translation [1]{[#1]}%
	\providecommand \BibitemOpen [0]{}%
	\providecommand \bibitemStop [0]{}%
	\providecommand \bibitemNoStop [0]{.\EOS\space}%
	\providecommand \EOS [0]{\spacefactor3000\relax}%
	\providecommand \BibitemShut  [1]{\csname bibitem#1\endcsname}%
	\let\auto@bib@innerbib\@empty
	\bibitem [{\citenamefont {Liz{\'a}rraga}\ \emph {et~al.}(2017)\citenamefont
		{Liz{\'a}rraga}, \citenamefont {Pan}, \citenamefont {Bergqvist},
		\citenamefont {Holmstr{\"o}m}, \citenamefont {Gercsi},\ and\ \citenamefont
		{Vitos}}]{2017_ScRep_Co}%
	\BibitemOpen
	\bibfield  {author} {\bibinfo {author} {\bibfnamefont {R.}~\bibnamefont
			{Liz{\'a}rraga}}, \bibinfo {author} {\bibfnamefont {F.}~\bibnamefont {Pan}},
		\bibinfo {author} {\bibfnamefont {L.}~\bibnamefont {Bergqvist}}, \bibinfo
		{author} {\bibfnamefont {E.}~\bibnamefont {Holmstr{\"o}m}}, \bibinfo {author}
		{\bibfnamefont {Z.}~\bibnamefont {Gercsi}},\ and\ \bibinfo {author}
		{\bibfnamefont {L.}~\bibnamefont {Vitos}},\ }\href@noop {} {\bibfield
		{journal} {\bibinfo  {journal} {Scientific reports}\ }\textbf {\bibinfo
			{volume} {7}},\ \bibinfo {pages} {1} (\bibinfo {year} {2017})}\BibitemShut {NoStop}%
	\bibitem [{\citenamefont {Matar}\ \emph {et~al.}(2007)\citenamefont {Matar},
		\citenamefont {Houari},\ and\ \citenamefont {Belkhir}}]{PRB:CoN:07}%
	\BibitemOpen
	\bibfield  {author} {\bibinfo {author} {\bibfnamefont {S.~F.}\ \bibnamefont
			{Matar}}, \bibinfo {author} {\bibfnamefont {A.}~\bibnamefont {Houari}},\ and\
		\bibinfo {author} {\bibfnamefont {M.~A.}\ \bibnamefont {Belkhir}},\
	}\href@noop {} {\bibfield  {journal} {\bibinfo  {journal} {Phys. Rev. B}\
		}\textbf {\bibinfo {volume} {75}},\ \bibinfo {pages} {245109} (\bibinfo
		{year} {2007})}\BibitemShut {NoStop}%
	\bibitem [{\citenamefont {Zeisberger}\ \emph {et~al.}(2007)\citenamefont
		{Zeisberger}, \citenamefont {Dutz}, \citenamefont {M{\"u}ller}, \citenamefont
		{Hergt}, \citenamefont {Matoussevitch},\ and\ \citenamefont
		{BÃ¶nnemann}}]{2007_JMMM_Co_application}%
	\BibitemOpen
	\bibfield  {author} {\bibinfo {author} {\bibfnamefont {M.}~\bibnamefont
			{Zeisberger}}, \bibinfo {author} {\bibfnamefont {S.}~\bibnamefont {Dutz}},
		\bibinfo {author} {\bibfnamefont {R.}~\bibnamefont {M{\"u}ller}}, \bibinfo
		{author} {\bibfnamefont {R.}~\bibnamefont {Hergt}}, \bibinfo {author}
		{\bibfnamefont {N.}~\bibnamefont {Matoussevitch}},\ and\ \bibinfo {author}
		{\bibfnamefont {H.}~\bibnamefont {B{\"a}nnemann}},\ }\href
	{https://doi.org/https://doi.org/10.1016/j.jmmm.2006.11.178} {\bibfield
		{journal} {\bibinfo  {journal} {Journal of Magnetism and Magnetic Materials}\
		}\textbf {\bibinfo {volume} {311}},\ \bibinfo {pages} {224} (\bibinfo {year}
		{2007})},\ 
	\BibitemShut
	{NoStop}%
	\bibitem [{\citenamefont {Boccato}\ \emph {et~al.}(2019)\citenamefont
		{Boccato}, \citenamefont {Torchio}, \citenamefont {D'Angelo}, \citenamefont
		{Trapananti}, \citenamefont {Kantor}, \citenamefont {Recoules}, \citenamefont
		{Anzellini}, \citenamefont {Morard}, \citenamefont {Irifune},\ and\
		\citenamefont {Pascarelli}}]{2019_PhysRevB_liq_Ni_Co_XAS}%
	\BibitemOpen
	\bibfield  {author} {\bibinfo {author} {\bibfnamefont {S.}~\bibnamefont
			{Boccato}}, \bibinfo {author} {\bibfnamefont {R.}~\bibnamefont {Torchio}},
		\bibinfo {author} {\bibfnamefont {P.}~\bibnamefont {D'Angelo}}, \bibinfo
		{author} {\bibfnamefont {A.}~\bibnamefont {Trapananti}}, \bibinfo {author}
		{\bibfnamefont {I.}~\bibnamefont {Kantor}}, \bibinfo {author} {\bibfnamefont
			{V.}~\bibnamefont {Recoules}}, \bibinfo {author} {\bibfnamefont
			{S.}~\bibnamefont {Anzellini}}, \bibinfo {author} {\bibfnamefont
			{G.}~\bibnamefont {Morard}}, \bibinfo {author} {\bibfnamefont
			{T.}~\bibnamefont {Irifune}},\ and\ \bibinfo {author} {\bibfnamefont
			{S.}~\bibnamefont {Pascarelli}},\ }\href@noop {} {\bibfield  {journal}
		{\bibinfo  {journal} {Phys. Rev. B}\ }\textbf {\bibinfo {volume} {100}},\
		\bibinfo {pages} {180101} (\bibinfo {year} {2019})}\BibitemShut {NoStop}%
	\bibitem [{\citenamefont {Torchio}\ \emph {et~al.}(2016)\citenamefont
		{Torchio}, \citenamefont {Marini}, \citenamefont {Kvashnin}, \citenamefont
		{Kantor}, \citenamefont {Mathon}, \citenamefont {Garbarino}, \citenamefont
		{Meneghini}, \citenamefont {Anzellini}, \citenamefont {Occelli},
		\citenamefont {Bruno}, \citenamefont {Dewaele},\ and\ \citenamefont
		{Pascarelli}}]{2016_PRB_Co_HPLT}%
	\BibitemOpen
	\bibfield  {author} {\bibinfo {author} {\bibfnamefont {R.}~\bibnamefont
			{Torchio}}, \bibinfo {author} {\bibfnamefont {C.}~\bibnamefont {Marini}},
		\bibinfo {author} {\bibfnamefont {Y.~O.}\ \bibnamefont {Kvashnin}}, \bibinfo
		{author} {\bibfnamefont {I.}~\bibnamefont {Kantor}}, \bibinfo {author}
		{\bibfnamefont {O.}~\bibnamefont {Mathon}}, \bibinfo {author} {\bibfnamefont
			{G.}~\bibnamefont {Garbarino}}, \bibinfo {author} {\bibfnamefont
			{C.}~\bibnamefont {Meneghini}}, \bibinfo {author} {\bibfnamefont
			{S.}~\bibnamefont {Anzellini}}, \bibinfo {author} {\bibfnamefont
			{F.}~\bibnamefont {Occelli}}, \bibinfo {author} {\bibfnamefont
			{P.}~\bibnamefont {Bruno}}, \bibinfo {author} {\bibfnamefont
			{A.}~\bibnamefont {Dewaele}},\ and\ \bibinfo {author} {\bibfnamefont
			{S.}~\bibnamefont {Pascarelli}},\ }\href
	{https://doi.org/10.1103/PhysRevB.94.024429} {\bibfield  {journal} {\bibinfo
			{journal} {Phys. Rev. B}\ }\textbf {\bibinfo {volume} {94}},\ \bibinfo
		{pages} {024429} (\bibinfo {year} {2016})}\BibitemShut {NoStop}%
	\bibitem [{\citenamefont {Yoo}\ \emph {et~al.}(1998)\citenamefont {Yoo},
		\citenamefont {S{\"a}derlind},\ and\ \citenamefont
		{Cynn}}]{1998_hcpfccphasediagram}%
	\BibitemOpen
	\bibfield  {author} {\bibinfo {author} {\bibfnamefont {C.-S.}\ \bibnamefont
			{Yoo}}, \bibinfo {author} {\bibfnamefont {P.}~\bibnamefont {S{\"a}derlind}},\
		and\ \bibinfo {author} {\bibfnamefont {H.}~\bibnamefont {Cynn}},\ }\href
	{https://doi.org/10.1088/0953-8984/10/20/001} {\bibfield  {journal} {\bibinfo
			{journal} {Journal of Physics: Condensed Matter}\ }\textbf {\bibinfo
			{volume} {10}},\ \bibinfo {pages} {L311} (\bibinfo {year}
		{1998})}\BibitemShut {NoStop}%
	\bibitem [{\citenamefont {Grass}\ and\ \citenamefont
		{Stark}(2006)}]{2006_GPSynthesis_Co}%
	\BibitemOpen
	\bibfield  {author} {\bibinfo {author} {\bibfnamefont {R.~N.}\ \bibnamefont
			{Grass}}\ and\ \bibinfo {author} {\bibfnamefont {W.~J.}\ \bibnamefont
			{Stark}},\ }\href {https://doi.org/10.1039/B601013J} {\bibfield  {journal}
		{\bibinfo  {journal} {J. Mater. Chem.}\ }\textbf {\bibinfo {volume} {16}},\
		\bibinfo {pages} {1825} (\bibinfo {year} {2006})}\BibitemShut {NoStop}%
	\bibitem [{\citenamefont {Zhang}\ \emph {et~al.}(2009)\citenamefont {Zhang},
		\citenamefont {Wang},\ and\ \citenamefont {Li}}]{2009_MCP_Co_synthesis}%
	\BibitemOpen
	\bibfield  {author} {\bibinfo {author} {\bibfnamefont {L.}~\bibnamefont
			{Zhang}}, \bibinfo {author} {\bibfnamefont {H.}~\bibnamefont {Wang}},\ and\
		\bibinfo {author} {\bibfnamefont {J.}~\bibnamefont {Li}},\ }\href
	{https://doi.org/https://doi.org/10.1016/j.matchemphys.2009.04.022}
	{\bibfield  {journal} {\bibinfo  {journal} {Materials Chemistry and Physics}\
		}\textbf {\bibinfo {volume} {116}},\ \bibinfo {pages} {514} (\bibinfo {year}
		{2009})}\BibitemShut {NoStop}%
	\bibitem [{\citenamefont {Gajbhiye}\ \emph {et~al.}(2008)\citenamefont
		{Gajbhiye}, \citenamefont {Sharma}, \citenamefont {Nigam},\ and\
		\citenamefont {Ningthoujam}}]{GAJBHIYE2008Co}%
	\BibitemOpen
	\bibfield  {author} {\bibinfo {author} {\bibfnamefont {N.~S.}\ \bibnamefont
			{Gajbhiye}}, \bibinfo {author} {\bibfnamefont {S.}~\bibnamefont {Sharma}},
		\bibinfo {author} {\bibfnamefont {A.~K.}\ \bibnamefont {Nigam}},\ and\
		\bibinfo {author} {\bibfnamefont {R.~S.}\ \bibnamefont {Ningthoujam}},\
	}\href {https://doi.org/https://doi.org/10.1016/j.cplett.2008.10.065}
	{\bibfield  {journal} {\bibinfo  {journal} {Chemical Physics Letters}\
		}\textbf {\bibinfo {volume} {466}},\ \bibinfo {pages} {181} (\bibinfo {year}
		{2008})}\BibitemShut {NoStop}%
	\bibitem [{\citenamefont {Kumar}\ and\ \citenamefont
		{Gupta}(2007)}]{2007_JMMM_DK}%
	\BibitemOpen
	\bibfield  {author} {\bibinfo {author} {\bibfnamefont {D.}~\bibnamefont
			{Kumar}}\ and\ \bibinfo {author} {\bibfnamefont {A.}~\bibnamefont {Gupta}},\
	}\href {https://doi.org/https://doi.org/10.1016/j.jmmm.2006.06.008}
	{\bibfield  {journal} {\bibinfo  {journal} {Journal of Magnetism and Magnetic
				Materials}\ }\textbf {\bibinfo {volume} {308}},\ \bibinfo {pages} {318}
		(\bibinfo {year} {2007})}\BibitemShut {NoStop}%
	\bibitem [{\citenamefont {Banu}\ \emph {et~al.}(2019)\citenamefont {Banu},
		\citenamefont {Kumar}, \citenamefont {Pandey}, \citenamefont {Satpati},
		\citenamefont {Gupta},\ and\ \citenamefont {Dev}}]{2019_TSF_Banu_Co}%
	\BibitemOpen
	\bibfield  {author} {\bibinfo {author} {\bibfnamefont {N.}~\bibnamefont
			{Banu}}, \bibinfo {author} {\bibfnamefont {P.}~\bibnamefont {Kumar}},
		\bibinfo {author} {\bibfnamefont {N.}~\bibnamefont {Pandey}}, \bibinfo
		{author} {\bibfnamefont {B.}~\bibnamefont {Satpati}}, \bibinfo {author}
		{\bibfnamefont {M.}~\bibnamefont {Gupta}},\ and\ \bibinfo {author}
		{\bibfnamefont {B.}~\bibnamefont {Dev}},\ }\href
	{https://doi.org/10.1016/j.tsf.2019.01.050} {\bibfield  {journal} {\bibinfo
			{journal} {Thin Solid Films}\ }\textbf {\bibinfo {volume} {675}},\ \bibinfo
		{pages} {177} (\bibinfo {year} {2019})}\BibitemShut {NoStop}%
	\bibitem [{\citenamefont {Andreev}\ \emph {et~al.}(2015)\citenamefont
		{Andreev}, \citenamefont {de~Lacaillerie}, \citenamefont {Lapina},\ and\
		\citenamefont {Gerashenko}}]{2015_hcp_fcc_Co_NMR}%
	\BibitemOpen
	\bibfield  {author} {\bibinfo {author} {\bibfnamefont {A.~S.}\ \bibnamefont
			{Andreev}}, \bibinfo {author} {\bibfnamefont {J.-B.~d.}\ \bibnamefont
			{de~Lacaillerie}}, \bibinfo {author} {\bibfnamefont {O.~B.}\ \bibnamefont
			{Lapina}},\ and\ \bibinfo {author} {\bibfnamefont {A.}~\bibnamefont
			{Gerashenko}},\ }\href@noop {} {\bibfield  {journal} {\bibinfo  {journal}
			{Physical Chemistry Chemical Physics}\ }\textbf {\bibinfo {volume} {17}},\
		\bibinfo {pages} {14598} (\bibinfo {year} {2015})}\BibitemShut {NoStop}%
	\bibitem [{\citenamefont {Kumar}\ \emph {et~al.}(2019)\citenamefont {Kumar},
		\citenamefont {Manjunatha}, \citenamefont {Anupama}, \citenamefont {Ramesh},\
		and\ \citenamefont {Sahoo}}]{2019_kumar_Co_in_C}%
	\BibitemOpen
	\bibfield  {author} {\bibinfo {author} {\bibfnamefont {R.}~\bibnamefont
			{Kumar}}, \bibinfo {author} {\bibfnamefont {M.}~\bibnamefont {Manjunatha}},
		\bibinfo {author} {\bibfnamefont {A.}~\bibnamefont {Anupama}}, \bibinfo
		{author} {\bibfnamefont {K.}~\bibnamefont {Ramesh}},\ and\ \bibinfo {author}
		{\bibfnamefont {B.}~\bibnamefont {Sahoo}},\ }\href
	{https://doi.org/https://doi.org/10.1016/j.ceramint.2019.06.243} {\bibfield
		{journal} {\bibinfo  {journal} {Ceramics International}\ }\textbf {\bibinfo
			{volume} {45}},\ \bibinfo {pages} {19879} (\bibinfo {year}
		{2019})}\BibitemShut {NoStop}%
	\bibitem [{\citenamefont {Banu}\ \emph {et~al.}(2017)\citenamefont {Banu},
		\citenamefont {Singh}, \citenamefont {Satpati}, \citenamefont {Roy},
		\citenamefont {Basu}, \citenamefont {Chakraborty}, \citenamefont {Movva},
		\citenamefont {Lauter},\ and\ \citenamefont {Dev}}]{2017_SR_Banu_HDNM_fccCo}%
	\BibitemOpen
	\bibfield  {author} {\bibinfo {author} {\bibfnamefont {N.}~\bibnamefont
			{Banu}}, \bibinfo {author} {\bibfnamefont {S.}~\bibnamefont {Singh}},
		\bibinfo {author} {\bibfnamefont {B.}~\bibnamefont {Satpati}}, \bibinfo
		{author} {\bibfnamefont {A.}~\bibnamefont {Roy}}, \bibinfo {author}
		{\bibfnamefont {S.}~\bibnamefont {Basu}}, \bibinfo {author} {\bibfnamefont
			{P.}~\bibnamefont {Chakraborty}}, \bibinfo {author} {\bibfnamefont
			{H.~C.~P.}\ \bibnamefont {Movva}}, \bibinfo {author} {\bibfnamefont
			{V.}~\bibnamefont {Lauter}},\ and\ \bibinfo {author} {\bibfnamefont {B.~N.}\
			\bibnamefont {Dev}},\ }\bibfield  {journal} {\bibinfo  {journal} {Scientific
			Reports}\ }\textbf {\bibinfo {volume} {7}},\ 
		(\bibinfo {year}
	{2017})\BibitemShut {NoStop}%
	\bibitem [{\citenamefont {Banu}\ \emph {et~al.}(2018)\citenamefont {Banu},
		\citenamefont {Singh}, \citenamefont {Basu}, \citenamefont {Roy},
		\citenamefont {Movva}, \citenamefont {Lauter}, \citenamefont {Satpati},\ and\
		\citenamefont {Dev}}]{2018_IOP_Banu_fccCo}%
	\BibitemOpen
	\bibfield  {author} {\bibinfo {author} {\bibfnamefont {N.}~\bibnamefont
			{Banu}}, \bibinfo {author} {\bibfnamefont {S.}~\bibnamefont {Singh}},
		\bibinfo {author} {\bibfnamefont {S.}~\bibnamefont {Basu}}, \bibinfo {author}
		{\bibfnamefont {A.}~\bibnamefont {Roy}}, \bibinfo {author} {\bibfnamefont
			{H.~C.~P.}\ \bibnamefont {Movva}}, \bibinfo {author} {\bibfnamefont
			{V.}~\bibnamefont {Lauter}}, \bibinfo {author} {\bibfnamefont
			{B.}~\bibnamefont {Satpati}},\ and\ \bibinfo {author} {\bibfnamefont {B.~N.}\
			\bibnamefont {Dev}},\ }\href {https://doi.org/10.1088/1361-6528/aab0e9}
	{\bibfield  {journal} {\bibinfo  {journal} {Nanotechnology}\ }\textbf
		{\bibinfo {volume} {29}},\ \bibinfo {pages} {195703} (\bibinfo {year}
		{2018})}\BibitemShut {NoStop}%
	\bibitem [{\citenamefont {Ohtake}\ \emph {et~al.}(2011)\citenamefont {Ohtake},
		\citenamefont {Yabuhara}, \citenamefont {Higuchi},\ and\ \citenamefont
		{Futamoto}}]{2011_ohtake_Co}%
	\BibitemOpen
	\bibfield  {author} {\bibinfo {author} {\bibfnamefont {M.~.}\ \bibnamefont
			{Ohtake}}, \bibinfo {author} {\bibfnamefont {O.~.}\ \bibnamefont {Yabuhara}},
		\bibinfo {author} {\bibfnamefont {J.~.}\ \bibnamefont {Higuchi}},\ and\
		\bibinfo {author} {\bibfnamefont {M.~.}\ \bibnamefont {Futamoto}},\ }\href
	{https://doi.org/10.1063/1.3537817} {\bibfield  {journal} {\bibinfo
			{journal} {Journal of Applied Physics}\ }\textbf {\bibinfo {volume} {109}},\
		\bibinfo {pages} {07C105} (\bibinfo {year} {2011})}\BibitemShut {NoStop}%
	\bibitem [{\citenamefont {Longo}\ \emph {et~al.}(2014)\citenamefont {Longo},
		\citenamefont {Sciortino}, \citenamefont {Giannici},\ and\ \citenamefont
		{Martorana}}]{longo2014_Co}%
	\BibitemOpen
	\bibfield  {author} {\bibinfo {author} {\bibfnamefont {A.}~\bibnamefont
			{Longo}}, \bibinfo {author} {\bibfnamefont {L.}~\bibnamefont {Sciortino}},
		\bibinfo {author} {\bibfnamefont {F.}~\bibnamefont {Giannici}},\ and\
		\bibinfo {author} {\bibfnamefont {A.}~\bibnamefont {Martorana}},\ }\href@noop
	{} {\bibfield  {journal} {\bibinfo  {journal} {Journal of Applied
				Crystallography}\ }\textbf {\bibinfo {volume} {47}},\ \bibinfo {pages} {1562}
		(\bibinfo {year} {2014})}\BibitemShut {NoStop}%
	\bibitem [{\citenamefont {Seema}\ \emph {et~al.}(2021)\citenamefont {Seema},
		\citenamefont {Tayal}, \citenamefont {Amir}, \citenamefont {P{\"u}tter},
		\citenamefont {Mattauch},\ and\ \citenamefont {Gupta}}]{2021_JAC_Seema_Co4N}%
	\BibitemOpen
	\bibfield  {author} {\bibinfo {author} {\bibnamefont {Seema}}, \bibinfo
		{author} {\bibfnamefont {A.}~\bibnamefont {Tayal}}, \bibinfo {author}
		{\bibfnamefont {S.}~\bibnamefont {Amir}}, \bibinfo {author} {\bibfnamefont
			{S.}~\bibnamefont {P{\"u}ter}}, \bibinfo {author} {\bibfnamefont
			{S.}~\bibnamefont {Mattauch}},\ and\ \bibinfo {author} {\bibfnamefont
			{M.}~\bibnamefont {Gupta}},\ }\href
	{https://doi.org/https://doi.org/10.1016/j.jallcom.2020.158052} {\bibfield
		{journal} {\bibinfo  {journal} {Journal of Alloys and Compounds}\ }\textbf
		{\bibinfo {volume} {863}},\ \bibinfo {pages} {158052} (\bibinfo {year}
		{2021})}\BibitemShut {NoStop}%
	\bibitem [{\citenamefont {Tol{\'e}dano}\ \emph {et~al.}(2001)\citenamefont
		{Tol{\'e}dano}, \citenamefont {Krexner}, \citenamefont {Prem}, \citenamefont
		{Weber},\ and\ \citenamefont {Dmitriev}}]{fcc_hcp_theory}%
	\BibitemOpen
	\bibfield  {author} {\bibinfo {author} {\bibfnamefont {P.}~\bibnamefont
			{Tol{\'e}dano}}, \bibinfo {author} {\bibfnamefont {G.}~\bibnamefont
			{Krexner}}, \bibinfo {author} {\bibfnamefont {M.}~\bibnamefont {Prem}},
		\bibinfo {author} {\bibfnamefont {H.-P.}\ \bibnamefont {Weber}},\ and\
		\bibinfo {author} {\bibfnamefont {V.}~\bibnamefont {Dmitriev}},\ }\href@noop
	{} {\bibfield  {journal} {\bibinfo  {journal} {Physical Review B}\ }\textbf
		{\bibinfo {volume} {64}},\ \bibinfo {pages} {144104} (\bibinfo {year}
		{2001})}\BibitemShut {NoStop}%
	\bibitem [{\citenamefont {Kang}\ \emph {et~al.}(2018)\citenamefont {Kang},
		\citenamefont {Kim}, \citenamefont {Yoon}, \citenamefont {Kim}, \citenamefont
		{Yang}, \citenamefont {Chung}, \citenamefont {Kim}, \citenamefont {Jeong},
		\citenamefont {Son}, \citenamefont {Kim} \emph {et~al.}}]{kang2018}%
	\BibitemOpen
	\bibfield  {author} {\bibinfo {author} {\bibfnamefont {J.~S.}\ \bibnamefont
			{Kang}}, \bibinfo {author} {\bibfnamefont {J.-Y.}\ \bibnamefont {Kim}},
		\bibinfo {author} {\bibfnamefont {J.}~\bibnamefont {Yoon}}, \bibinfo {author}
		{\bibfnamefont {J.}~\bibnamefont {Kim}}, \bibinfo {author} {\bibfnamefont
			{J.}~\bibnamefont {Yang}}, \bibinfo {author} {\bibfnamefont {D.~Y.}\
			\bibnamefont {Chung}}, \bibinfo {author} {\bibfnamefont {M.-c.}\ \bibnamefont
			{Kim}}, \bibinfo {author} {\bibfnamefont {H.}~\bibnamefont {Jeong}}, \bibinfo
		{author} {\bibfnamefont {Y.~J.}\ \bibnamefont {Son}}, \bibinfo {author}
		{\bibfnamefont {B.~G.}\ \bibnamefont {Kim}}, \emph {et~al.},\ }\href@noop {}
	{\bibfield  {journal} {\bibinfo  {journal} {Advanced Energy Materials}\
		}\textbf {\bibinfo {volume} {8}},\ \bibinfo {pages} {1703114} (\bibinfo
		{year} {2018})}\BibitemShut {NoStop}%
	\bibitem [{\citenamefont {Pandey}\ \emph {et~al.}(2017)\citenamefont {Pandey},
		\citenamefont {Gupta}, \citenamefont {Gupta}, \citenamefont {Chakravarty},
		\citenamefont {Shukla},\ and\ \citenamefont {Devishvili}}]{JAC16_NPandey}%
	\BibitemOpen
	\bibfield  {author} {\bibinfo {author} {\bibfnamefont {N.}~\bibnamefont
			{Pandey}}, \bibinfo {author} {\bibfnamefont {M.}~\bibnamefont {Gupta}},
		\bibinfo {author} {\bibfnamefont {R.}~\bibnamefont {Gupta}}, \bibinfo
		{author} {\bibfnamefont {S.}~\bibnamefont {Chakravarty}}, \bibinfo {author}
		{\bibfnamefont {N.}~\bibnamefont {Shukla}},\ and\ \bibinfo {author}
		{\bibfnamefont {A.}~\bibnamefont {Devishvili}},\ }\href@noop {} {\bibfield
		{journal} {\bibinfo  {journal} {Journal of Alloys and Compounds}\ }\textbf
		{\bibinfo {volume} {694}},\ \bibinfo {pages} {1209 } (\bibinfo {year}
		{2017})}\BibitemShut {NoStop}%
	\bibitem [{\citenamefont {Fang}\ \emph {et~al.}(2004)\citenamefont {Fang},
		\citenamefont {Yang}, \citenamefont {Hsu}, \citenamefont {Chen},
		\citenamefont {Lin},\ and\ \citenamefont {Chen}}]{JVSTA:Fang:CoN}%
	\BibitemOpen
	\bibfield  {author} {\bibinfo {author} {\bibfnamefont {J.-S.}\ \bibnamefont
			{Fang}}, \bibinfo {author} {\bibfnamefont {L.-C.}\ \bibnamefont {Yang}},
		\bibinfo {author} {\bibfnamefont {C.-S.}\ \bibnamefont {Hsu}}, \bibinfo
		{author} {\bibfnamefont {G.-S.}\ \bibnamefont {Chen}}, \bibinfo {author}
		{\bibfnamefont {Y.-W.}\ \bibnamefont {Lin}},\ and\ \bibinfo {author}
		{\bibfnamefont {G.-S.}\ \bibnamefont {Chen}},\ }\href@noop {} {\bibfield
		{journal} {\bibinfo  {journal} {Journal of Vacuum Science and Technology A}\
		}\textbf {\bibinfo {volume} {22}} (\bibinfo {year} {2004})}\BibitemShut
	{NoStop}%
	\bibitem [{\citenamefont {Gupta}\ \emph {et~al.}(2015)\citenamefont {Gupta},
		\citenamefont {Pandey}, \citenamefont {Tayal},\ and\ \citenamefont
		{Gupta}}]{CoN_AIP_Adv2015}%
	\BibitemOpen
	\bibfield  {author} {\bibinfo {author} {\bibfnamefont {R.}~\bibnamefont
			{Gupta}}, \bibinfo {author} {\bibfnamefont {N.}~\bibnamefont {Pandey}},
		\bibinfo {author} {\bibfnamefont {A.}~\bibnamefont {Tayal}},\ and\ \bibinfo
		{author} {\bibfnamefont {M.}~\bibnamefont {Gupta}},\ }\href
	{https://doi.org/https://doi.org/10.1063/1.4930977} {\bibfield  {journal}
		{\bibinfo  {journal} {AIP Advances}\ }\textbf {\bibinfo {volume} {5}},\
		\bibinfo {eid} {097131} (\bibinfo {year} {2015})}\BibitemShut {NoStop}%
	\bibitem [{\citenamefont {Tayal}\ \emph
		{et~al.}(2014{\natexlab{a}})\citenamefont {Tayal}, \citenamefont {Gupta},
		\citenamefont {Lalla}, \citenamefont {Gupta}, \citenamefont {Horisberger},
		\citenamefont {Stahn}, \citenamefont {Schlage},\ and\ \citenamefont
		{Wille}}]{PRB:AT:2014}%
	\BibitemOpen
	\bibfield  {author} {\bibinfo {author} {\bibfnamefont {A.}~\bibnamefont
			{Tayal}}, \bibinfo {author} {\bibfnamefont {M.}~\bibnamefont {Gupta}},
		\bibinfo {author} {\bibfnamefont {N.~P.}\ \bibnamefont {Lalla}}, \bibinfo
		{author} {\bibfnamefont {A.}~\bibnamefont {Gupta}}, \bibinfo {author}
		{\bibfnamefont {M.}~\bibnamefont {Horisberger}}, \bibinfo {author}
		{\bibfnamefont {J.}~\bibnamefont {Stahn}}, \bibinfo {author} {\bibfnamefont
			{K.}~\bibnamefont {Schlage}},\ and\ \bibinfo {author} {\bibfnamefont {H.-C.}\
			\bibnamefont {Wille}},\ }\href@noop {} {\bibfield  {journal} {\bibinfo
			{journal} {Phys. Rev. B}\ }\textbf {\bibinfo {volume} {90}},\ \bibinfo
		{pages} {144412} (\bibinfo {year} {2014}{\natexlab{a}})}\BibitemShut
	{NoStop}%
	\bibitem [{\citenamefont {Pandey}\ \emph
		{et~al.}(2019{\natexlab{a}})\citenamefont {Pandey}, \citenamefont {Gupta},
		\citenamefont {Gupta}, \citenamefont {Hussain}, \citenamefont {Reddy},
		\citenamefont {Phase},\ and\ \citenamefont {Stahn}}]{PhysRevB.NP}%
	\BibitemOpen
	\bibfield  {author} {\bibinfo {author} {\bibfnamefont {N.}~\bibnamefont
			{Pandey}}, \bibinfo {author} {\bibfnamefont {M.}~\bibnamefont {Gupta}},
		\bibinfo {author} {\bibfnamefont {R.}~\bibnamefont {Gupta}}, \bibinfo
		{author} {\bibfnamefont {Z.}~\bibnamefont {Hussain}}, \bibinfo {author}
		{\bibfnamefont {V.~R.}\ \bibnamefont {Reddy}}, \bibinfo {author}
		{\bibfnamefont {D.~M.}\ \bibnamefont {Phase}},\ and\ \bibinfo {author}
		{\bibfnamefont {J.}~\bibnamefont {Stahn}},\ }\href@noop {} {\bibfield
		{journal} {\bibinfo  {journal} {Phys. Rev. B}\ }\textbf {\bibinfo {volume}
			{99}},\ \bibinfo {pages} {214109} (\bibinfo {year}
		{2019}{\natexlab{a}})}\BibitemShut {NoStop}%
	\bibitem [{\citenamefont {Imai}\ \emph {et~al.}(2014)\citenamefont {Imai},
		\citenamefont {Sohma},\ and\ \citenamefont {Suemasu}}]{imai2014}%
	\BibitemOpen
	\bibfield  {author} {\bibinfo {author} {\bibfnamefont {Y.}~\bibnamefont
			{Imai}}, \bibinfo {author} {\bibfnamefont {M.}~\bibnamefont {Sohma}},\ and\
		\bibinfo {author} {\bibfnamefont {T.}~\bibnamefont {Suemasu}},\ }\href@noop
	{} {\bibfield  {journal} {\bibinfo  {journal} {Journal of alloys and
				compounds}\ }\textbf {\bibinfo {volume} {611}},\ \bibinfo {pages} {440}
		(\bibinfo {year} {2014})}\BibitemShut {NoStop}%
	\bibitem [{\citenamefont {van Straaten}\ \emph {et~al.}(2020)\citenamefont {van
			Straaten}, \citenamefont {Deckers}, \citenamefont {Vos}, \citenamefont
		{Kessels},\ and\ \citenamefont {Creatore}}]{2020_acs_jpcc_CoNx}%
	\BibitemOpen
	\bibfield  {author} {\bibinfo {author} {\bibfnamefont {G.}~\bibnamefont {van
				Straaten}}, \bibinfo {author} {\bibfnamefont {R.}~\bibnamefont {Deckers}},
		\bibinfo {author} {\bibfnamefont {M.~F.~J.}\ \bibnamefont {Vos}}, \bibinfo
		{author} {\bibfnamefont {W.~M.~M.}\ \bibnamefont {Kessels}},\ and\ \bibinfo
		{author} {\bibfnamefont {M.}~\bibnamefont {Creatore}},\ }\href
	{https://doi.org/10.1021/acs.jpcc.0c04223} {\bibfield  {journal} {\bibinfo
			{journal} {The Journal of Physical Chemistry C}\ }\textbf {\bibinfo {volume}
			{124}},\ \bibinfo {pages} {22046} (\bibinfo {year} {2020})}\BibitemShut
	{NoStop}%
	\bibitem [{\citenamefont {Phase}\ \emph {et~al.}(2014)\citenamefont {Phase},
		\citenamefont {Gupta}, \citenamefont {Potdar}, \citenamefont {Behera},
		\citenamefont {Sah},\ and\ \citenamefont {Gupta}}]{XAS_beamline}%
	\BibitemOpen
	\bibfield  {author} {\bibinfo {author} {\bibfnamefont {D.~M.}\ \bibnamefont
			{Phase}}, \bibinfo {author} {\bibfnamefont {M.}~\bibnamefont {Gupta}},
		\bibinfo {author} {\bibfnamefont {S.}~\bibnamefont {Potdar}}, \bibinfo
		{author} {\bibfnamefont {L.}~\bibnamefont {Behera}}, \bibinfo {author}
		{\bibfnamefont {R.}~\bibnamefont {Sah}},\ and\ \bibinfo {author}
		{\bibfnamefont {A.}~\bibnamefont {Gupta}},\ }\href@noop {} {\bibfield
		{journal} {\bibinfo  {journal} {AIP Conference Proceedings}\ }\textbf
		{\bibinfo {volume} {1591}},\ \bibinfo {pages} {685} (\bibinfo {year}
		{2014})}\BibitemShut {NoStop}%
	\bibitem [{\citenamefont {Asahara}\ \emph {et~al.}(2001)\citenamefont
		{Asahara}, \citenamefont {Migita}, \citenamefont {Tanaka},\ and\
		\citenamefont {Kawabata}}]{2001_Vac_Asahara}%
	\BibitemOpen
	\bibfield  {author} {\bibinfo {author} {\bibfnamefont {H.}~\bibnamefont
			{Asahara}}, \bibinfo {author} {\bibfnamefont {T.}~\bibnamefont {Migita}},
		\bibinfo {author} {\bibfnamefont {T.}~\bibnamefont {Tanaka}},\ and\ \bibinfo
		{author} {\bibfnamefont {K.}~\bibnamefont {Kawabata}},\ }\href
	{https://doi.org/https://doi.org/10.1016/S0042-207X(00)00453-X} {\bibfield
		{journal} {\bibinfo  {journal} {Vacuum}\ }\textbf {\bibinfo {volume} {62}},\
		\bibinfo {pages} {293 } (\bibinfo {year} {2001})}\BibitemShut {NoStop}%
	\bibitem [{\citenamefont {Holzwarth}\ and\ \citenamefont
		{Gibson}(2011)}]{Scherrer_Eq:NatureNano:2011}%
	\BibitemOpen
	\bibfield  {author} {\bibinfo {author} {\bibfnamefont {U.}~\bibnamefont
			{Holzwarth}}\ and\ \bibinfo {author} {\bibfnamefont {N.}~\bibnamefont
			{Gibson}},\ }\href@noop {} {\bibfield  {journal} {\bibinfo  {journal} {Nature
				Nano.}\ }\textbf {\bibinfo {volume} {6}},\ \bibinfo {pages} {534} (\bibinfo
		{year} {2011})}\BibitemShut {NoStop}%
	\bibitem [{\citenamefont {Gupta}\ \emph {et~al.}(2019)\citenamefont {Gupta},
		\citenamefont {Seema}, \citenamefont {Pandey}, \citenamefont {Amir},
		\citenamefont {Putter},\ and\ \citenamefont {Mattauch}}]{JMMM_MG}%
	\BibitemOpen
	\bibfield  {author} {\bibinfo {author} {\bibfnamefont {M.}~\bibnamefont
			{Gupta}}, \bibinfo {author} {\bibnamefont {Seema}}, \bibinfo {author}
		{\bibfnamefont {N.}~\bibnamefont {Pandey}}, \bibinfo {author} {\bibfnamefont
			{S.}~\bibnamefont {Amir}}, \bibinfo {author} {\bibfnamefont {S.}~\bibnamefont
			{Putter}},\ and\ \bibinfo {author} {\bibfnamefont {S.}~\bibnamefont
			{Mattauch}},\ }\href@noop {} {\bibfield  {journal} {\bibinfo  {journal}
			{Journal of Magnetism and Magnetic Materials}\ }\textbf {\bibinfo {volume}
			{489}},\ \bibinfo {pages} {165376} (\bibinfo {year} {2019})}\BibitemShut
	{NoStop}%
	\bibitem [{\citenamefont {Beshkova}\ \emph {et~al.}(2001)\citenamefont
		{Beshkova}, \citenamefont {Beshkov}, \citenamefont {Marinov}, \citenamefont
		{Bogdanov-Dimitrov}, \citenamefont {Mladenov}, \citenamefont {Tanaka},\ and\
		\citenamefont {Kawabata}}]{RTA_CoxN}%
	\BibitemOpen
	\bibfield  {author} {\bibinfo {author} {\bibfnamefont {M.}~\bibnamefont
			{Beshkova}}, \bibinfo {author} {\bibfnamefont {G.}~\bibnamefont {Beshkov}},
		\bibinfo {author} {\bibfnamefont {M.}~\bibnamefont {Marinov}}, \bibinfo
		{author} {\bibfnamefont {D.}~\bibnamefont {Bogdanov-Dimitrov}}, \bibinfo
		{author} {\bibfnamefont {G.}~\bibnamefont {Mladenov}}, \bibinfo {author}
		{\bibfnamefont {T.}~\bibnamefont {Tanaka}},\ and\ \bibinfo {author}
		{\bibfnamefont {K.}~\bibnamefont {Kawabata}},\ }\href
	{https://doi.org/10.1081/AMP-100108525} {\bibfield  {journal} {\bibinfo
			{journal} {Materials and Manufacturing Processes}\ }\textbf {\bibinfo
			{volume} {16}},\ \bibinfo {pages} {531} (\bibinfo {year} {2001})},
		\BibitemShut {NoStop}%
	\bibitem [{\citenamefont {Pandey}\ \emph
		{et~al.}(2019{\natexlab{b}})\citenamefont {Pandey}, \citenamefont {Gupta},
		\citenamefont {Rawat}, \citenamefont {Amir}, \citenamefont {Stahn},\ and\
		\citenamefont {Gupta}}]{2019_physB_Fe4N_NP}%
	\BibitemOpen
	\bibfield  {author} {\bibinfo {author} {\bibfnamefont {N.}~\bibnamefont
			{Pandey}}, \bibinfo {author} {\bibfnamefont {M.}~\bibnamefont {Gupta}},
		\bibinfo {author} {\bibfnamefont {R.}~\bibnamefont {Rawat}}, \bibinfo
		{author} {\bibfnamefont {S.}~\bibnamefont {Amir}}, \bibinfo {author}
		{\bibfnamefont {J.}~\bibnamefont {Stahn}},\ and\ \bibinfo {author}
		{\bibfnamefont {A.}~\bibnamefont {Gupta}},\ }\href@noop {} {\bibfield
		{journal} {\bibinfo  {journal} {Physica B: Condensed Matter}\ }\textbf
		{\bibinfo {volume} {572}},\ \bibinfo {pages} {36} (\bibinfo {year}
		{2019}{\natexlab{b}})}\BibitemShut {NoStop}%
	\bibitem [{\citenamefont {Pandey}\ \emph
		{et~al.}(2019{\natexlab{c}})\citenamefont {Pandey}, \citenamefont {Gupta},
		\citenamefont {Gupta}, \citenamefont {Amir},\ and\ \citenamefont
		{Stahn}}]{2019NP_Co_1_3inch}%
	\BibitemOpen
	\bibfield  {author} {\bibinfo {author} {\bibfnamefont {N.}~\bibnamefont
			{Pandey}}, \bibinfo {author} {\bibfnamefont {M.}~\bibnamefont {Gupta}},
		\bibinfo {author} {\bibfnamefont {R.}~\bibnamefont {Gupta}}, \bibinfo
		{author} {\bibfnamefont {S.~M.}\ \bibnamefont {Amir}},\ and\ \bibinfo
		{author} {\bibfnamefont {J.}~\bibnamefont {Stahn}},\ }\href
	{https://doi.org/https://doi.org/10.1007/s00339-019-2825-0} {\bibfield
		{journal} {\bibinfo  {journal} {Applied Physics A}\ }\textbf {\bibinfo
			{volume} {125}},\ \bibinfo {pages} {539} (\bibinfo {year}
		{2019}{\natexlab{c}})}\BibitemShut {NoStop}%
	\bibitem [{\citenamefont {Yao}\ \emph {et~al.}(2019)\citenamefont {Yao},
		\citenamefont {Li}, \citenamefont {Zhou}, \citenamefont {Zhao}, \citenamefont
		{Cheng}, \citenamefont {Chen},\ and\ \citenamefont {Luo}}]{yao2019CrCo4N}%
	\BibitemOpen
	\bibfield  {author} {\bibinfo {author} {\bibfnamefont {N.}~\bibnamefont
			{Yao}}, \bibinfo {author} {\bibfnamefont {P.}~\bibnamefont {Li}}, \bibinfo
		{author} {\bibfnamefont {Z.}~\bibnamefont {Zhou}}, \bibinfo {author}
		{\bibfnamefont {Y.}~\bibnamefont {Zhao}}, \bibinfo {author} {\bibfnamefont
			{G.}~\bibnamefont {Cheng}}, \bibinfo {author} {\bibfnamefont
			{S.}~\bibnamefont {Chen}},\ and\ \bibinfo {author} {\bibfnamefont
			{W.}~\bibnamefont {Luo}},\ }\href@noop {} {\bibfield  {journal} {\bibinfo
			{journal} {Advanced Energy Materials}\ }\textbf {\bibinfo {volume} {9}},\
		\bibinfo {pages} {1902449} (\bibinfo {year} {2019})}\BibitemShut {NoStop}%
	\bibitem [{\citenamefont {Kim}\ \emph {et~al.}(2006)\citenamefont {Kim},
		\citenamefont {Lee}, \citenamefont {Park},\ and\ \citenamefont
		{Ahn}}]{kim2006CoSi2}%
	\BibitemOpen
	\bibfield  {author} {\bibinfo {author} {\bibfnamefont {S.~I.}\ \bibnamefont
			{Kim}}, \bibinfo {author} {\bibfnamefont {S.~R.}\ \bibnamefont {Lee}},
		\bibinfo {author} {\bibfnamefont {J.~H.}\ \bibnamefont {Park}},\ and\
		\bibinfo {author} {\bibfnamefont {B.~T.}\ \bibnamefont {Ahn}},\ }\href@noop
	{} {\bibfield  {journal} {\bibinfo  {journal} {Journal of The Electrochemical
				Society}\ }\textbf {\bibinfo {volume} {153}},\ \bibinfo {pages} {G506}
		(\bibinfo {year} {2006})}\BibitemShut {NoStop}%
	\bibitem [{\citenamefont {Guo}\ \emph {et~al.}(2018)\citenamefont {Guo},
		\citenamefont {Wang}, \citenamefont {Li}, \citenamefont {Yang}, \citenamefont
		{Tamirat}, \citenamefont {Qi}, \citenamefont {Han}, \citenamefont {Li},
		\citenamefont {Wang},\ and\ \citenamefont {Feng}}]{2018CoCo4NMAB}%
	\BibitemOpen
	\bibfield  {author} {\bibinfo {author} {\bibfnamefont {Z.}~\bibnamefont
			{Guo}}, \bibinfo {author} {\bibfnamefont {F.}~\bibnamefont {Wang}}, \bibinfo
		{author} {\bibfnamefont {Z.}~\bibnamefont {Li}}, \bibinfo {author}
		{\bibfnamefont {Y.}~\bibnamefont {Yang}}, \bibinfo {author} {\bibfnamefont
			{A.~G.}\ \bibnamefont {Tamirat}}, \bibinfo {author} {\bibfnamefont
			{H.}~\bibnamefont {Qi}}, \bibinfo {author} {\bibfnamefont {J.}~\bibnamefont
			{Han}}, \bibinfo {author} {\bibfnamefont {W.}~\bibnamefont {Li}}, \bibinfo
		{author} {\bibfnamefont {L.}~\bibnamefont {Wang}},\ and\ \bibinfo {author}
		{\bibfnamefont {S.}~\bibnamefont {Feng}},\ }\href@noop {} {\bibfield
		{journal} {\bibinfo  {journal} {Journal of Materials Chemistry A}\ }\textbf
		{\bibinfo {volume} {6}},\ \bibinfo {pages} {22096} (\bibinfo {year}
		{2018})}\BibitemShut {NoStop}%
	\bibitem [{\citenamefont {Fan}\ \emph {et~al.}(2019)\citenamefont {Fan},
		\citenamefont {Sang}, \citenamefont {Jiang}, \citenamefont {Yang},
		\citenamefont {Zhang}, \citenamefont {Chen},\ and\ \citenamefont
		{Liu}}]{fan2019ALD}%
	\BibitemOpen
	\bibfield  {author} {\bibinfo {author} {\bibfnamefont {Q.}~\bibnamefont
			{Fan}}, \bibinfo {author} {\bibfnamefont {L.}~\bibnamefont {Sang}}, \bibinfo
		{author} {\bibfnamefont {D.}~\bibnamefont {Jiang}}, \bibinfo {author}
		{\bibfnamefont {L.}~\bibnamefont {Yang}}, \bibinfo {author} {\bibfnamefont
			{H.}~\bibnamefont {Zhang}}, \bibinfo {author} {\bibfnamefont
			{Q.}~\bibnamefont {Chen}},\ and\ \bibinfo {author} {\bibfnamefont
			{Z.}~\bibnamefont {Liu}},\ }\href@noop {} {\bibfield  {journal} {\bibinfo
			{journal} {Journal of Vacuum Science \& Technology A: Vacuum, Surfaces, and
				Films}\ }\textbf {\bibinfo {volume} {37}},\ \bibinfo {pages} {010904}
		(\bibinfo {year} {2019})}\BibitemShut {NoStop}%
	\bibitem [{\citenamefont {Guo}\ \emph {et~al.}(2020)\citenamefont {Guo},
		\citenamefont {Wan}, \citenamefont {Li}, \citenamefont {Xi},\ and\
		\citenamefont {Wang}}]{2020_AFM_TiNCo4N}%
	\BibitemOpen
	\bibfield  {author} {\bibinfo {author} {\bibfnamefont {D.}~\bibnamefont
			{Guo}}, \bibinfo {author} {\bibfnamefont {Z.}~\bibnamefont {Wan}}, \bibinfo
		{author} {\bibfnamefont {Y.}~\bibnamefont {Li}}, \bibinfo {author}
		{\bibfnamefont {B.}~\bibnamefont {Xi}},\ and\ \bibinfo {author}
		{\bibfnamefont {C.}~\bibnamefont {Wang}},\ }\href
	{https://doi.org/https://doi.org/10.1002/adfm.202008511} {\bibfield
		{journal} {\bibinfo  {journal} {Advanced Functional Materials}\ }\textbf
		{\bibinfo {volume} {31}},\ \bibinfo {pages} {2008511} (\bibinfo {year}
		{2020})},
	 \BibitemShut
	{NoStop}%
	\bibitem [{\citenamefont {Jiao}\ \emph {et~al.}()\citenamefont {Jiao},
		\citenamefont {Wang}, \citenamefont {Chen}, \citenamefont {Zhang},
		\citenamefont {Mou}, \citenamefont {Zhang},\ and\ \citenamefont
		{Liu}}]{Co4N_CNT}%
	\BibitemOpen
	\bibfield  {author} {\bibinfo {author} {\bibfnamefont {M.}~\bibnamefont
			{Jiao}}, \bibinfo {author} {\bibfnamefont {Z.}~\bibnamefont {Wang}}, \bibinfo
		{author} {\bibfnamefont {Z.}~\bibnamefont {Chen}}, \bibinfo {author}
		{\bibfnamefont {X.}~\bibnamefont {Zhang}}, \bibinfo {author} {\bibfnamefont
			{K.}~\bibnamefont {Mou}}, \bibinfo {author} {\bibfnamefont {W.}~\bibnamefont
			{Zhang}},\ and\ \bibinfo {author} {\bibfnamefont {L.}~\bibnamefont {Liu}},\
	}\href {https://doi.org/https://doi.org/10.1002/celc.202000062} {\bibfield
		{journal} {\bibinfo  {journal} {ChemElectroChem}\ }\textbf {\bibinfo {volume}
			{7}},\ \bibinfo {pages} {2065}},\ \Eprint
	{https://arxiv.org/abs/https://chemistry-europe.onlinelibrary.wiley.com/doi/pdf/10.1002/celc.202000062}
	{https://chemistry-europe.onlinelibrary.wiley.com/doi/pdf/10.1002/celc.202000062}
	\BibitemShut {NoStop}%
	\bibitem [{\citenamefont {Yao}\ \emph {et~al.}(2007)\citenamefont {Yao},
		\citenamefont {Zhu}, \citenamefont {Chen}, \citenamefont {Wang},
		\citenamefont {Au},\ and\ \citenamefont {Shi}}]{2007_Yao_Co4N_Al2O3}%
	\BibitemOpen
	\bibfield  {author} {\bibinfo {author} {\bibfnamefont {Z.}~\bibnamefont
			{Yao}}, \bibinfo {author} {\bibfnamefont {A.}~\bibnamefont {Zhu}}, \bibinfo
		{author} {\bibfnamefont {J.}~\bibnamefont {Chen}}, \bibinfo {author}
		{\bibfnamefont {X.}~\bibnamefont {Wang}}, \bibinfo {author} {\bibfnamefont
			{C.}~\bibnamefont {Au}},\ and\ \bibinfo {author} {\bibfnamefont
			{C.}~\bibnamefont {Shi}},\ }\href@noop {} {\bibfield  {journal} {\bibinfo
			{journal} {Journal of Solid State Chemistry}\ }\textbf {\bibinfo {volume}
			{180}},\ \bibinfo {pages} {2635} (\bibinfo {year} {2007})}\BibitemShut
	{NoStop}%
	\bibitem [{\citenamefont {Li}\ \emph {et~al.}(2017)\citenamefont {Li},
		\citenamefont {Zhang}, \citenamefont {Yang}, \citenamefont {Liu},
		\citenamefont {Zhu},\ and\ \citenamefont {Zhou}}]{li2017CoN}%
	\BibitemOpen
	\bibfield  {author} {\bibinfo {author} {\bibfnamefont {H.}~\bibnamefont
			{Li}}, \bibinfo {author} {\bibfnamefont {Y.}~\bibnamefont {Zhang}}, \bibinfo
		{author} {\bibfnamefont {K.}~\bibnamefont {Yang}}, \bibinfo {author}
		{\bibfnamefont {H.}~\bibnamefont {Liu}}, \bibinfo {author} {\bibfnamefont
			{X.}~\bibnamefont {Zhu}},\ and\ \bibinfo {author} {\bibfnamefont
			{H.}~\bibnamefont {Zhou}},\ }\href@noop {} {\bibfield  {journal} {\bibinfo
			{journal} {Applied Surface Science}\ }\textbf {\bibinfo {volume} {406}},\
		\bibinfo {pages} {110} (\bibinfo {year} {2017})}\BibitemShut {NoStop}%
	\bibitem [{\citenamefont {Shi}\ and\ \citenamefont
		{Lederman}(2000)}]{2000annealedCo_MA}%
	\BibitemOpen
	\bibfield  {author} {\bibinfo {author} {\bibfnamefont {H.}~\bibnamefont
			{Shi}}\ and\ \bibinfo {author} {\bibfnamefont {D.}~\bibnamefont {Lederman}},\
	}\href@noop {} {\bibfield  {journal} {\bibinfo  {journal} {Journal of Applied
				Physics}\ }\textbf {\bibinfo {volume} {87}},\ \bibinfo {pages} {6095}
		(\bibinfo {year} {2000})}\BibitemShut {NoStop}%
	\bibitem [{\citenamefont {Matsuoka}\ \emph {et~al.}(1986)\citenamefont
		{Matsuoka}, \citenamefont {Ono},\ and\ \citenamefont
		{Inukai}}]{1996_APL_Co-N_matsouka}%
	\BibitemOpen
	\bibfield  {author} {\bibinfo {author} {\bibfnamefont {M.}~\bibnamefont
			{Matsuoka}}, \bibinfo {author} {\bibfnamefont {K.}~\bibnamefont {Ono}},\ and\
		\bibinfo {author} {\bibfnamefont {T.}~\bibnamefont {Inukai}},\ }\href@noop {}
	{\bibfield  {journal} {\bibinfo  {journal} {Applied physics letters}\
		}\textbf {\bibinfo {volume} {49}},\ \bibinfo {pages} {977} (\bibinfo {year}
		{1986})}\BibitemShut {NoStop}%
	\bibitem [{\citenamefont {Tayal}\ \emph
		{et~al.}(2014{\natexlab{b}})\citenamefont {Tayal}, \citenamefont {Gupta},
		\citenamefont {Kumar}, \citenamefont {Reddy}, \citenamefont {Gupta},
		\citenamefont {Amir}, \citenamefont {Korelis},\ and\ \citenamefont
		{Stahn}}]{JAP_AT_FeN_Al_Zr}%
	\BibitemOpen
	\bibfield  {author} {\bibinfo {author} {\bibfnamefont {A.}~\bibnamefont
			{Tayal}}, \bibinfo {author} {\bibfnamefont {M.}~\bibnamefont {Gupta}},
		\bibinfo {author} {\bibfnamefont {D.}~\bibnamefont {Kumar}}, \bibinfo
		{author} {\bibfnamefont {V.~R.}\ \bibnamefont {Reddy}}, \bibinfo {author}
		{\bibfnamefont {A.}~\bibnamefont {Gupta}}, \bibinfo {author} {\bibfnamefont
			{S.~M.}\ \bibnamefont {Amir}}, \bibinfo {author} {\bibfnamefont
			{P.}~\bibnamefont {Korelis}},\ and\ \bibinfo {author} {\bibfnamefont
			{J.}~\bibnamefont {Stahn}},\ }\href@noop {} {\bibfield  {journal} {\bibinfo
			{journal} {Journal of Applied Physics}\ }\textbf {\bibinfo {volume} {116}},\
		\bibinfo {pages} {222206} (\bibinfo {year} {2014}{\natexlab{b}})}\BibitemShut
	{NoStop}%
	\bibitem [{\citenamefont {Shewmon}(1963)}]{Shewmon}%
	\BibitemOpen
	\bibfield  {author} {\bibinfo {author} {\bibfnamefont {P.~G.}\ \bibnamefont
			{Shewmon}},\ }\href@noop {} {\emph {\bibinfo {title} {Diffusion in Solids}}}\
	(\bibinfo  {publisher} {Mc Graw-Hill, New York},\ \bibinfo {year}
	{1963})\BibitemShut {NoStop}%
	\bibitem [{\citenamefont {Gupta}(2020)}]{MG_chap}%
	\BibitemOpen
	\bibfield  {author} {\bibinfo {author} {\bibfnamefont {M.}~\bibnamefont
			{Gupta}},\ }\bibinfo {title} {Synthesis, stability and self-diffusion in iron
		nitride thin films: A review},\ in\ \href@noop {} {\emph {\bibinfo
			{booktitle} {Recent Advances in Thin Films}}},\ \bibinfo {editor} {edited by\
		\bibinfo {editor} {\bibfnamefont {S.}~\bibnamefont {Kumar}}\ and\ \bibinfo
		{editor} {\bibfnamefont {D.~K.}\ \bibnamefont {Aswal}}}\ (\bibinfo
	{publisher} {Springer Singapore},\ 
	\ \bibinfo
	{year} {2020})\ pp.\ \bibinfo {pages} {131--179}\BibitemShut {NoStop}%
	\bibitem [{\citenamefont {Pandey}\ \emph {et~al.}(2020)\citenamefont {Pandey},
		\citenamefont {Gupta},\ and\ \citenamefont {Stahn}}]{2020_RRL_NiN_NP}%
	\BibitemOpen
	\bibfield  {author} {\bibinfo {author} {\bibfnamefont {N.}~\bibnamefont
			{Pandey}}, \bibinfo {author} {\bibfnamefont {M.}~\bibnamefont {Gupta}},\ and\
		\bibinfo {author} {\bibfnamefont {J.}~\bibnamefont {Stahn}},\ }\href
	{https://doi.org/https://doi.org/10.1002/pssr.202000294} {\bibfield
		{journal} {\bibinfo  {journal} {physica status solidi (RRL) Rapid
				Research Letters}\ }\textbf {\bibinfo {volume} {14}},\ \bibinfo {pages}
		{2000294} (\bibinfo {year} {2020})},
\BibitemShut
	{NoStop}%
	\bibitem [{\citenamefont {Chatbi}\ \emph {et~al.}(1997)\citenamefont {Chatbi},
		\citenamefont {Vergnat}, \citenamefont {Bobo},\ and\ \citenamefont
		{Hennet}}]{chatbi1997nitrogen}%
	\BibitemOpen
	\bibfield  {author} {\bibinfo {author} {\bibfnamefont {H.}~\bibnamefont
			{Chatbi}}, \bibinfo {author} {\bibfnamefont {M.}~\bibnamefont {Vergnat}},
		\bibinfo {author} {\bibfnamefont {J.-F.}\ \bibnamefont {Bobo}},\ and\
		\bibinfo {author} {\bibfnamefont {L.}~\bibnamefont {Hennet}},\ }\href@noop {}
	{\bibfield  {journal} {\bibinfo  {journal} {Solid state communications}\
		}\textbf {\bibinfo {volume} {102}},\ \bibinfo {pages} {677} (\bibinfo {year}
		{1997})}\BibitemShut {NoStop}%
\end{thebibliography}

%

\end{document}